\documentclass[aps,prd,twocolumn,groupedaddress,nofootinbib]{revtex4-1}

\usepackage{graphicx,color,amsmath}
\usepackage[utf8]{inputenc}

\newcommand{\second}{{\, {\rm s}}}

\newcommand{\metre}{{\, {\rm m}}}
\newcommand{\cm}{{\, {\rm cm}}}

\newcommand{\eV}{{\, {\rm eV}}}

\newcommand{\GeV}{{\, {\rm GeV}}}

\newcommand{\kelvin}{{\, {\rm K}}}

\newcommand{\Hz}{{\, {\rm Hz}}}
\newcommand{\kHz}{{\, {\rm kHz}}}
\newcommand{\MHz}{{\, {\rm MHz}}}
\newcommand{\GHz}{{\, {\rm GHz}}}

\newcommand{\LL}{{\mathcal L}}
\newcommand{\OO}{{\mathcal O}}

\newcommand{\Tesla}{{\, {\rm T}}}
\newcommand{\Joule}{{\, {\rm J}}}
\newcommand{\Watt}{{\, {\rm W}}}
\newcommand{\litre}{{\, {\rm L}}}
\newcommand{\kW}{{\, {\rm kW}}}
\newcommand{\kJ}{{\, {\rm kJ}}}

\newcommand{\TM}{{\rm TM}}
\newcommand{\TE}{{\rm TE}}

\begin{document}

\title{Microwave cavity searches for low-frequency axion dark matter}

\author{Robert Lasenby}
\email{rlasenby@stanford.edu}
\affiliation{Stanford Institute for Theoretical Physics, Stanford University, Stanford, CA 94305, USA}

\date{\today}

\begin{abstract}
	For low-mass (frequency $\ll \GHz$) axions, dark matter
	detection experiments searching for an axion-photon-photon
	coupling generally have suppressed sensitivity, 
	if they use a static background magnetic field.
	This geometric suppression can be alleviated by using a
	high-frequency oscillating background field.
	Here, we present a high-level sketch
	of such an experiment, using superconducting cavities
	at $\sim \GHz$ frequencies. We discuss the physical limits
	on signal power arising from cavity properties,
	and point out cavity geometries that could
	circumvent some of these limitations.
	We also consider how backgrounds, including
	vibrational noise and drive signal leakage,
	might impact sensitivity. While practical microwave
	field strengths are significantly below attainable
	static magnetic fields, the lack of geometric
	suppression, and higher quality factors, may allow
	superconducting cavity experiments to be competitive
	in some regimes.
\end{abstract}

\maketitle

\tableofcontents


\section{Introduction}

While dark matter (DM) has so far only been observed
through its gravitational interactions, many
theoretical candidates have additional interactions 
with the Standard Model (SM). This has motivated a wide
range of laboratory searches for DM, most famously through
the WIMP direct detection program.
Another well-motivated dark matter candidate is
the QCD axion (and more generally, axion-like particles),
for which an extensive experimental
program also exists~\cite{Graham:2015ouw}.

A major part of this program
involves searches for axion
DM through the $a F_{\mu\nu} \tilde F^{\mu\nu}$
axion-photon-photon coupling.
This coupling is a generic 
prediction of QCD axion models~\cite{diCortona:2015ldu,DiLuzio:2016sbl}, and arises in many other theories of axion-like
DM~\cite{Jaeckel:2010ni}.
For axion masses corresponding to 
frequencies $\gtrsim \GHz$, simple
experimental designs such as cavity
haloscopes (e.g.\ the ADMX experiment~\cite{Du:2018uak,Braine:2019fqb}) or dielectric haloscopes~\cite{TheMADMAXWorkingGroup:2016hpc,Baryakhtar:2018doz} can achieve almost-optimal~\cite{rnl}
axion-mass-averaged power absorption from the DM field.

However, the situation is different
for much smaller axion masses, corresponding
to Compton wavelengths significantly longer
than the length scale of the experiment
(either the shielding scale or the extent of the magnetic
field, whichever is smaller). 
The EM fields at these frequencies
are naturally in the quasi-static regime,
and for axion detection experiments using a static
background magnetic field, this means that
the absorbed power 
is parametrically suppressed by $\sim (m_a L)^2$,
where $L$ is the experimental length scale.
The origin of this suppression is discussed
in a number of papers~\cite{Ouellet:2018nfr,Beutter:2018xfx,rnl};
roughly speaking, it arises because the electric field oscillations
associated with current fluctuations are suppressed
compared to the magnetic field oscillations, reducing
the interaction with the axion-sourced
effective current.
This suppression affects low-frequency axion detection
proposals such as ABRACADABRA~\cite{Kahn:2016aff} and DM Radio~\cite{scshort}, reducing
their sensitivity.\footnote{While the sensitivity
of experiments such as ABRACADABRA is not directly related
to absorbed power, the $g_{a \gamma \gamma}$
sensitivity is still suppressed by $\sim (m_a L)$, compared
to its theoretical scaling at higher frequencies~\cite{rnl}.}

One way to avoid these issues is to use an oscillating
background magnetic field. If this is oscillating 
at a frequency $\omega_0$, then it combines with the
axion field oscillation to generate an effective
current oscillating at $\omega_0 \pm m_a$.
By taking $\omega_0 \gtrsim L^{-1} \gg m_a$, 
the response excited by this effective current
is no longer in the quasi-static regime, and
the absorbed power is no longer suppressed.
Since the signal excitation is at a much higher
frequency than the axion field,
this approach is referred to as `up-conversion'
\cite{Goryachev:2018vjt}.
The idea of using an oscillating background magnetic field
in axion detection experiments was first
proposed in~\cite{Sikivie:2010fa}, but
they were mostly concerned with GHz-scale axion frequencies,
and did not consider the parametric scaling at low axion masses.

Up-conversion experiments using optical-frequency
background fields have been proposed
in~\cite{DeRocco:2018jwe,Obata:2018vvr,Liu:2018icu}.
However, these encounter a number of issues.
The most serious is that achievable magnetic field
strengths at optical frequencies are very small,
compared to static magnetic fields. 
Another is that, even if other noise sources
are overcome, the shot noise suppression
coming from absorbing fewer, but higher-energy,
optical photons
degrades the theoretical sensitivity
limits still further. Consequently, it would be very difficult
to match the sensitivity of static-field experiments,
despite the lack of a $(m_a L)^2$ suppression.

We can address both of these issues by using lower-frequency magnetic
field oscillations. In particular, reasonably large magnetic
fields ($\sim 0.2 \Tesla$) are routinely attained at $\sim \GHz$
frequencies inside superconducting (SRF) cavities~\cite{padamsee1998rf,Grassellino:2017bod},
and as we discuss below, it may be possible to achieve
even higher fields. The lower frequency also
alleviates shot noise issues. 
Thus, while the average magnetic fields will still be lower
than those used in static-field experiments,
the relative $(m_a L)^2$ enhancement may be enough to make
up-conversion experiments interesting.

SRF up-conversion experiments were first
proposed in~\cite{Sikivie:2010fa}, but as
stated above, they were mainly concerned with
$\sim \GHz$ axion frequencies. Up-conversion
experiments for low-mass axions were considered
in~\cite{Goryachev:2018vjt}, but calculational
errors resulted in parametrically
incorrect sensitivity limits, which were orders of
magnitude too optimistic.\footnote{
	Specifically, the overlap term $\xi_-$ in e.g.\ 
	equation 20 of~\cite{Goryachev:2018vjt} should
	go to zero as their two modes become degenerate,
	which does not seem to occur in their calculations;
	for more details, see~\cite{rnl}.
	}

In this paper, we discuss the basic physics and design
considerations involved in microwave cavity up-conversion
experiments. One of the most obvious questions is how
to choose the cavity geometry. We derive constraints
on the signal power attainable from cavities,
and show that for simple geometries (in which all of the walls
are visible from an interior point) the RMS magnetic field
is limited by the magnetic field at the walls.
Since, for superconducting cavities, this this limited
by the superconductor's material properties, the signal
power from such cavities is bounded. We also show 
how this bound can be circumvented using
cavities with more complicated shapes, and illustrate
that, to probe significant QCD axion parameter space
with small cavity volumes, such
geometries may be practically necessary.

In addition to such considerations, which would be important
for more advanced experiments, we also outline a
nominal first-generation experiment, aiming to be as simple
as possible while still having interesting reach.
We give a high-level discussion of the noise issues
that might arise, and derive representative sensitivity 
estimates. This discussion is not intended as a design
study, and a realistic experiment might be significantly
more complicated. Instead, our goal is to illustrate 
the physical parameters that might be required, and to motivate further study of this experimental direction.


\section{Axion DM up-conversion}
\label{secupconv}

We will assume
that dark matter consists of a light axion-like particle $a$,
which couples to the SM via the electromagnetic
$F_{\mu\nu} \tilde F^{\mu\nu}$ operator.
This has Lagrangian\footnote{
We take the $(+---)$ signature, and use the convention
$\epsilon_{0123} = -1$. Except where indicated, we use natural units
with $c = \hbar = 1$.
In general, we will abbreviate $g_{a \gamma \gamma} = g$.}
\begin{align}
	\LL &\supset \frac{1}{2} (\partial_\mu a)^2 - V(a)
-\frac{1}{4} g_{a \gamma \gamma} a F_{\mu \nu} \tilde{F}^{\mu\nu} \nonumber\\
	&=\frac{1}{2} (\partial_\mu a)^2 - V(a) +g_{a \gamma \gamma} a E \cdot B,
\end{align}
where $V(a)$ is the potential for the axion --- in general,
only the mass term $V(a) = \frac{1}{2} m_a^2 a^2$ will be important
for us.

For light DM ($m_a \ll \eV$), the occupation number in
the Milky Way is $\gg 1$, and almost all cosmological
histories result in its state today being a coherent, classical-like
oscillation~\cite{Kawasaki:2014sqa,Ringwald:2015dsf,Co:2017mop,diCortona:2015ldu,Graham:2015rva}. Since $g_{a \gamma\gamma}$ (and other
couplings) are constrained to be very small, interactions
with a detector will have a negligible effect on the DM's state.
Consequently, for the purposes of
detection, we can treat the DM oscillation as a fixed classical
background field. 

Under integration by parts, the interaction term is equivalent
to $\LL \supset - \frac{1}{2} A_\mu J^\mu_{(a)}$,
where the `axion current' $J^\mu_{(a)}$ is given by
\begin{equation}
	J^\mu_{(a)} = g \begin{pmatrix} - \nabla a \cdot B \\
		\dot a B + \nabla a \times E
	\end{pmatrix}
\end{equation}
Since axion DM in the galaxy is non-relativistic, with
typical velocity $\sim 10^{-3}$, the dominant
term is the spatial current $J_{(a)} \simeq g \dot a B$.
Hence, the effects of the axion field are equivalent
to those of an oscillating current density, with profile
set by the background magnetic field.

As described in the introduction,
we will consider `up-conversion' experiments,
in which the background magnetic field
is oscillating at
a frequency $\omega_0 \gg m_a$.
Writing the background magnetic field
as
$B_0(t,x) \simeq B_0 \cos (\omega_0 t) b(x)$, and
given a single-frequency axion oscillation,
$a(t) = a_0 \cos (m_a t)$, the effective
current $J_{(a)}$ will have frequency components
at $\omega_0 \pm m_a$. 
As illustrated in Figure~\ref{figupconv},
the basic idea is to arrange things so
that there is another EM mode
with resonant frequency $\simeq \omega_0 + m_a$ (or $\omega_0 - m_a$),
so that it can efficiently absorb power from the
effective current oscillation.
The larger this absorbed power, the easier it
is to detect an axion DM signal (other things being equal).

For an EM mode with electric field profile
$E_1(x)$, the instantaneous power input from
the axion current is $P_a = - \int dV
 E_1 \cdot J_{(a)}$. Considering only a single frequency component
 of $J_{(a)}$, say $\omega_J = m_a + \omega_B$,
the cycle-averaged input power at
this frequency is
is
\begin{equation}
\bar P_a = \frac{1}{4} \cos\alpha \, g \, \omega_J a_0 B_0 \int dV E_1 \cdot b
\end{equation}
where $\alpha$ is the relative phase of
the electric field response and the $-J_{(a)}$ oscillation.
The average power dissipated is $P_{\rm diss}
= \omega_J U / Q_l$, where 
$U = \frac{1}{2} \int dV E_1^2$ is
the stored energy in the mode (at this frequency),
and $Q_l$ is the (loaded) quality factor of 
the mode. In a steady state, 
the absorbed and dissipated powers will be equal.
Equating $\bar P_a = P_{\rm diss}$, we find that the cycle-averaged
absorbed power, once fully rung up, is 
\begin{equation}
	P_{\rm sig} = \frac{1}{8} \cos^2\alpha (g a_0 B_0)^2 m_a^2 \frac{Q_l}{\omega_J}
	\frac{\left(\int dV b \cdot E_1 \right)^2}{\int dV E_1^2}
\end{equation}
We can define the geometric overlap factor
\begin{equation}
	C_{01} \equiv \frac{\left(\int dV b \cdot E_1\right)^2}
	{\left( \int dV E_1^2 \right) \left(\int dV b^2 \right)}
\end{equation}
which measures the degree of overlap between
the background magnetic field and the
$E_1$ electric field; this has
$C_{01} \le 1$, with equality iff $E_1(x) \propto b(x)$.
Using this, we can write the signal power as
\begin{equation}
	P_{\rm sig}
	= \frac{1}{4} \cos^2 \alpha (g a_0)^2 m_a^2 
	\frac{Q_l}{\omega_J} C_{01} U_0
	\label{eqp1}
\end{equation}
where $U_0 = \frac{1}{2} B_0^2 V_b$.
For a high quality factor mode,
if the signal frequency $\omega$ is close to the resonance
frequency $\omega_1$, then 
\begin{equation}
	\cos^2\alpha \simeq \frac{1}{1 + Q_l^2 \frac{(\omega^2 - \omega_1^2)^2}{\omega_1^4}}
	\simeq \frac{1}{1 + 4 Q_l^2 \left(\frac{\omega}{\omega_1} - 1\right)^2}
\end{equation}

The expression in equation~\ref{eqp1} is equivalent to equation 3 in~\cite{Sikivie:2010fa}, and gives the cycle-averaged
power absorbed by a particular mode
(assuming that the mode bandwidth is small
compared to $m_a$).
In most of the cases we are interested in, only
one mode of the cavity will have a resonance
frequency close to the signal frequency, 
so only that mode will absorb appreciable power from the axion
current.

In many circumstances (see appendix~\ref{apsnr}), we 
are interested in the signal power averaged 
over different axion masses. If we integrate over an
axion mass range $\Delta m$ significantly larger
than the bandwidth of the target mode, then the
average power absorbed is
\begin{equation}
	\bar P \simeq \frac{\pi}{4} g^2 \frac{\rho}{\Delta m} C_{01} U_0
	\label{eqmapow}
\end{equation}
This formula is valid even for low-frequency $B_0$
oscillations, as long as the integration time is long enough
to resolve the variation of $B$. However, if $\omega_0 \ll
L^{-1}$, where $L^{-1}$ is the linear scale of the shielded
experimental volume (or the magnetic field extent, whichever
is smaller), then $C_{01}$ is generally suppressed compared
to the theoretical limit, with $C_{01} \sim (\omega_0 L)^2$.
This is because the EM fields in the volume are
in the quasi-static regime, as discussed in~\cite{rnl}.
Consequently, to avoid this geometrical suppression, we want
to take $\omega_0 \gtrsim L^{-1}$, i.e.\ $\gtrsim \GHz$ for
laboratory-scale experiments. 
In such cases, if $C_{01}$ is close to 1, then 
the absorbed power can be close to the theoretically
obtainable limit, for a given background
magnetic field energy~\cite{rnl}.

A common limitation on the rate at which large magnetic fields
can be varied is the large amount of field energy stored
--- taking some nominal parameters, ${\rm Tesla}^2 \times \metre^3 \sim {\rm MJ}$. Feeding that amount of energy into 
and out of a system must generally be done rather
slowly. For example, the currents through high-field superconducting 
magnets generally take minutes or more to build
up to their full values, with faster changes
damaging the system. 
Faster rates of change are possible with
specially-designed superconducting systems~\cite{tinkham1996introduction},
but the dissipated power increases with frequency,
and is generally prohibitive for
rates of change $\gtrsim 1 \Tesla / {\rm s}$.
Resistive conductors can tolerate more heating
/ stresses, but sustained operation at high
field strengths dissipates very large powers.
Similarly, varying the fields from magnetic materials
will either involve mechanical motion, or hysteretic
energy losses. In all of these cases, achieving strong
$\sim \GHz$ frequency magnetic fields will not be possible.

\subsection{SRF cavities}
\label{secsrf}

We can get around these issues by using cavities
with high quality factors,
in which magnetic field energy
is exchanged back and forth with the electric field
energy inside the cavity, rather than needing to be
transferred in and out each cycle.
Filling the cavity to high field amplitude is still a
slow process, but once this amplitude has been established,
a high-field oscillation can be maintained with only a small
energy input, to counteract the small dissipation rate.

\begin{figure*}[t]
	\begin{center}
		\includegraphics[width=.9\linewidth]{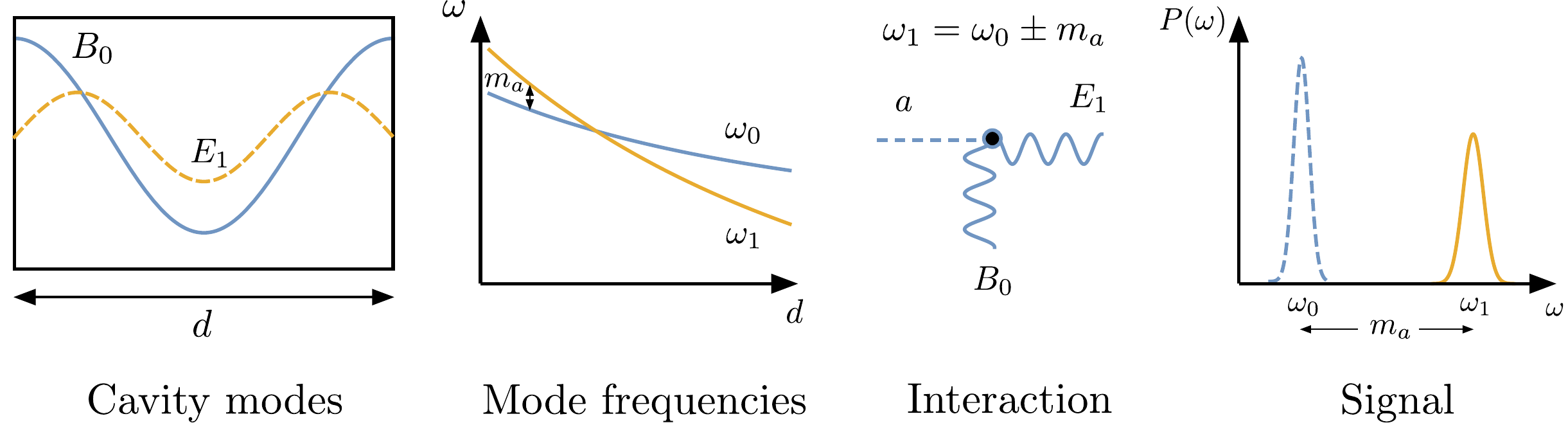}
		\caption{Schematic illustration of an up-conversion
		experiment for axion DM detection.
  The left-most panel (`cavity modes')
  illustrates a cavity in which we
  are interested in two modes, one
  with magnetic field profile $B_0$,
  and another with electric field
  profile $E_1$. The adjacent graph
  (`mode frequencies') illustrates the
  frequencies of these modes as we vary
  the shape of the cavity (here, the
  length $d$) , showing that they become
  degenerate at some $d$. When the
  cavity length is tuned to be close to
  this point, the frequency splitting
  between the modes will be small. If
  the $\omega_0$ mode is driven with
  high amplitude,
  then in the presence of an axion DM
oscillation with $m_a \simeq |\omega_1
- \omega_0|$, power will be transferred
		to the $\omega_1$ mode (illustrated
		in the `interaction' panel).
		This signal can be detected as 
		excess power, above thermal (or quantum) fluctuations,
		in the $\omega_1$ mode, as indicated
		in the `signal' panel.
		While some of the power in the $\omega_0$ mode
		may leak into the detected signal
		(indicated as a dashed peak in the `signal' panel),
		the frequency difference between $\omega_0$ and $\omega_1$
		helps to distinguish this from the axion signal (see
		Section~\ref{secleakage}).
		The mode profiles shown here are schematic --- Figure~\ref{figCylModes1} shows actual mode profiles for a particular
		cavity geometry.
		}
	\label{figupconv}
	\end{center}
\end{figure*}

The basic setup of a cavity up-conversion experiment
is illustrated in figure~\ref{figupconv}.
The cavity's shape is tuned so that there are two
modes with frequency difference $|\omega_1 - \omega_0| \simeq m_a$.
One of these modes is driven to a high field amplitude,
and in the presence of this background oscillation,
an axion DM oscillation would lead to signal power in the other
mode, according to equation~\ref{eqp1}.
To maximise the signal power, we want $C_{01}$ to
be close to one.

To achieve a high quality factor, and allow for high
drive fields without excessive energy dissipation,
the cavity walls should be superconducting.
SRF (superconducting radio frequency) cavities
have been extensively developed for
particle acceleration~\cite{padamsee1998rf}.
They are also starting to be used directly
in hidden sector
particle searches --- the Fermilab DarkSRF project~\cite{darksrf}
is constructing a dark photon detection experiment
using SRF cavities, and there have been
other proposals for axion detection~\cite{Bogorad:2019pbu,Janish:2019dpr}.
For axion DM detection experiments,
oscillating background magnetic fields
inside SRF cavities were first proposed in~\cite{Sikivie:2010fa}.
In this section, we will review some of the important
properties of SRF cavities, which will affect the
design and operation of an up-conversion experiment.

\begin{itemize}
	\item \emph{Peak surface magnetic field}: if the
		magnetic field at the walls of the cavity
		becomes too large, the behaviour of the superconducting
		material will change.
		Type-I superconductors
		generally have rather low critical fields
		(e.g. for Aluminium, $H_c \simeq 0.01 \Tesla$~\cite{Cochran_1958}),
		so are unsuitable for high-field cavities.
		SRF cavities are fabricated using Type-II
		superconductors, and almost always use niobium. This has a critical
		field value $H_{c_1}$ above which vortices
		penetrate; if this happens, then the radio-frequency
		oscillation of these vortices will lead to increased
		dissipation, and generally runaway thermal instability~\cite{padamsee1998rf}.
		It is actually possible to operate
		slightly above $H_{c_1}$, in a metastable
		`Meissner state' --- while it energetically favourable
		to have flux deep within the bulk, establishing this
		configuration involves penetrating a surface
		energy barrier~\cite{padamsee1998rf}. Vortices start penetrating
		the surface at $H_{sh}$, which for niobium
		is $\simeq 0.2 \Tesla$. Consequently, the magnetic
		field at the cavity walls should always be $\lesssim 0.2 \Tesla$. As we will see below, this puts limits on the achievable
		fields inside the cavity, and correspondingly,
		on the signal power attainable from axion DM.

	\item \emph{Surface resistance}: the quality factor of
		a cavity mode is set by its magnetic fields
		at the cavity walls (which determine 
		the wall currents), and the surface resistance there.
		This resistance has a `BCS' component,
		and a `residual' component,
		\begin{equation}
			R_s = R_{\rm BCS} + R_{\rm res}
		\end{equation}
		The BCS component, for oscillations
		at frequency $\omega$, is
		given approximately by
		\begin{equation}
			R_{\rm BCS}\simeq \omega^2 \lambda^3 \sigma_n \frac{\Delta}{T}
			\log\left(\frac{2.246 T}{\omega}\right)
			e^{-\Delta /T}
		\end{equation}
		where $\lambda$ is the (effective) penetration depth, 
		$\sigma_n$ is the normal-state conductivity, and $\Delta$ 
		is the superconducting gap energy~\cite{Gurevich:2012vua}.
		The residual resistance is the component that
		persists as $T \rightarrow 0$,
		and is measured to be $\sim {\rm few \, n\Omega}$
		for good niobium cavities (physically,
		it is not entirely clear what this residual resistance
		is dominated by~\cite{padamsee1998rf,Gurevich:2012vua}).
		The $R_{\rm BCS} \propto \omega^2$ dependence means
		that, for frequencies
		$\gtrsim 3 \GHz$, the BCS resistance starts to dominate.
		Since it is an increasing function of temperature,
		this can lead to thermal instability problems~\cite{padamsee1998rf}.
		Hence, SRF cavities are generally operated at lower frequencies.

	\item \emph{Cooling:} The quality factor of a mode sets the power
		dissipated, for a given energy stored. As we will 
		we see in the next section, taking the $H_{sh}$
		limit and $R_s$ values from above, and applying
		them to a simple laboratory-scale ($\sim 60 \litre$) cavity,
		gives $P_{\rm diss} \sim 30 \Watt$ (with $Q \simeq 2 \times 10^{11}$).
		Since this heat eventually needs
		to be dissipated to the $T_H \sim 300 \kelvin$ environment, 
		the maximum efficiency of the cooling system
		is $\eta_C = \frac{T_0}{T_H - T_0} \simeq T_0/T_H \simeq
		3 \times 10^{-3} T_0/\kelvin$, where $T_0$ is the
		temperature of the cavity. So, we would need at least
		$10 \kW$ of electrical power to cool the cavity to $T_0 = 1\kelvin$. Taking a typical thermal efficiency of $\eta_T \simeq 0.2$, this becomes $\gtrsim 50 \kW$. These figures illustrate that cooling the cavity to significantly sub-kelvin temperature would
		be prohibitively power-hungry. The high cooling
		powers required generally necessitate the use of
		liquid helium cooling systems. The pump machinery
		involved leads to mechanical vibrations of the cavity,
		which can introduce noise and tuning issues,
		as we discuss in section~\ref{secvibs}.

	\item \emph{Field emission:} if the electric fields at
		the cavity walls are high enough, then electrons
		can escape from the surface via tunneling.
		The field emission rate can be approximated
		via a modified Fowler-Nordheim formula~\cite{padamsee1998rf},
		\begin{equation}
			I_F \simeq 10^{-7} \frac{\phi}{\eV}
			\frac{A_e (\beta E_0)^2}{\phi^2}
			\exp\left(-6 \phi^{3/2} \sqrt{m_e}/(\beta E_0)\right) 
		\end{equation}
		where $\phi$ is the work function 
		of the wall material, $A_e$ is the effective emitting
		area, $E_0$ is the electric field at the wall,
		and $\beta$ is a phenomenological `field enhancement
		factor'. 
		For pure niobium, $\phi \simeq 4.3 \eV$~\cite{Michaelson_1977}, so 
		$6 \phi^{3/2} \sqrt{m_e} \simeq 60 {\rm \, GV} /\metre
		\simeq 200 \Tesla$. As we will see in
		section~\ref{seccavgeom}, for the cavities
		of interest to us, the peak electric field at the
		walls will be $\lesssim$ the peak magnetic
		field, which is restricted to $H_{sh} \simeq 0.2 \Tesla$.
		Consequently, if $\beta \sim 1$, we would naively expect field
		emission to be negligible.
		However, experimentally, field emission 
		is observed at significantly lower electric fields,
		down to $\sim 10 {\rm MV} / \metre \simeq 0.03 \Tesla$
		\cite{Padamsee:1998ya}. This appears
		to arise from defects (especially foreign objects,
		such as metallic fragments) on the walls of the cavity,
		which can have $\beta$ up to $\sim 700$~\cite{Padamsee:1998ya}.
		Too much field emission can lead to quality factor
		degradation, as EM energy is lost to electrons.
		Even if this is not a concern,
		we may be worried about much smaller levels
		of field emission, if the electrons deposit energy
		into the signal mode.
		We discuss this noise source in section~\ref{secfreecharge}.

	\item \emph{Tuning}: to change the axion mass that we are
		sensitive to, we need to change the frequency
		splitting between the drive and signal modes,
		which entails changing the shape of the cavity.
		If the axion mass is small compared
		to the mode frequencies,
		then only a small fractional change
		in the frequency of each mode is needed to
		cover an $\OO(1)$ axion mass range, simplifying
		the tuning problem.
		The usual method of tuning SRF cavities is simply 
		via elastic deformation of the cavity walls,
		through an external forcing (such as a piezoelectric
		transducer, and/or a mechanical screw)~\cite{padamsee1998rf,Pischalnikov:2015eye}.
		To stay within the (low-temperature) elastic
		limit of niobium, the material strain
		should be $\lesssim {\rm few} \times 10^{-3}$~\cite{Antoine:2011zz}.
		For typical $\sim \GHz$ cavities, this usually
		translates to a tunable range of a few hundred kHZ
		\cite{padamsee1998rf} --- for higher cavity modes,
		the absolute range may be greater.
		As well as static deformations,
		we also need to worry about vibrations, as
		discussed in section~\ref{secvibs}.
\end{itemize}

These points only represent a simple sketch
of the issues that an experiment would encounter,
and much more detailed analysis would be needed
for a real implementation. However, they give
some idea of the properties
of SRF cavities that are relevant to our setup.

\subsection{Cavity geometry}
\label{seccavgeom}

\begin{figure*}[t]
	\begin{center}
		\includegraphics[width=.8\linewidth]{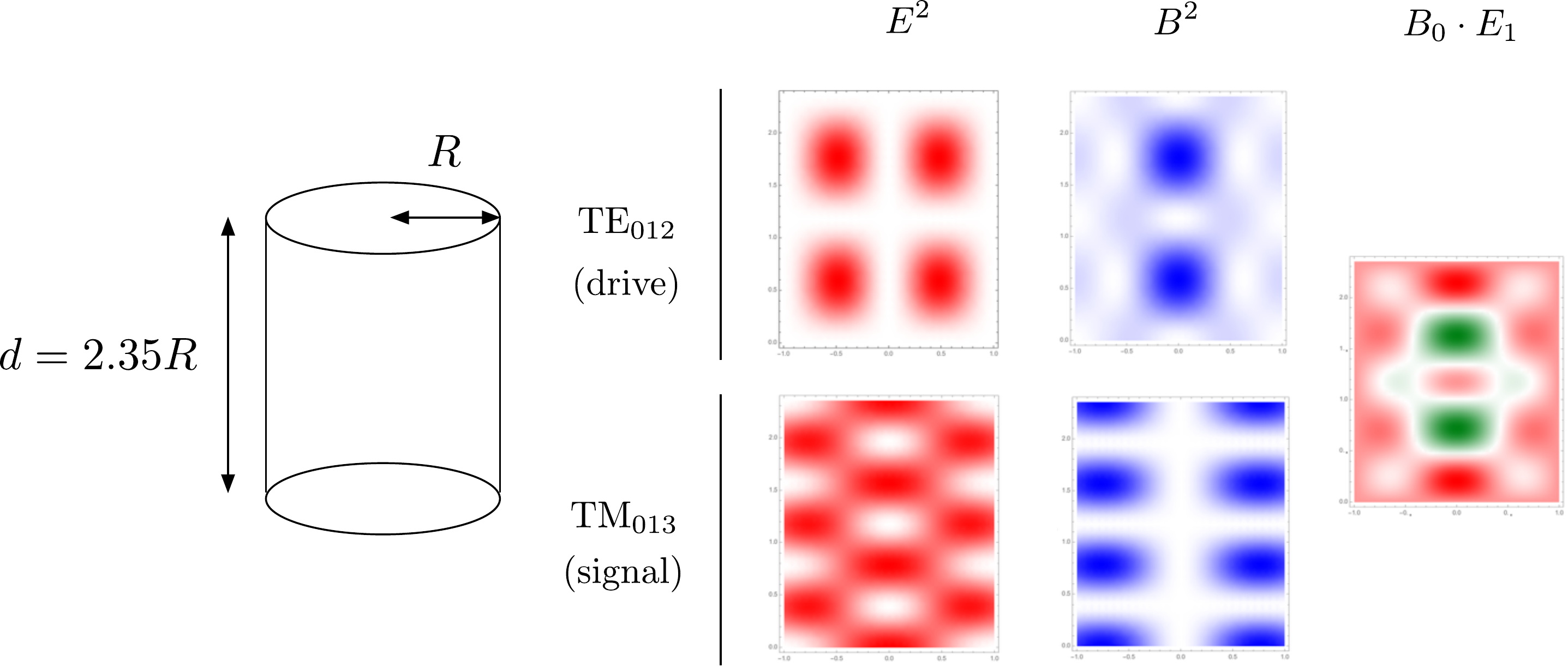}
		\caption{Illustration of the optimum drive and signal
		modes for a cylindrical cavity, as discussed in
		section~\ref{seccavgeom}.
		The $B_0 \cdot E_1$ panel shows the dot product of the drive
		mode's magnetic field with the signal mode's electric field, with
		green indicating a positive value, and red a negative value
		--- the integrated value over the volume gives $C_{01} \simeq 0.19$, as defined in section~\ref{secupconv}.}
	\label{figCylModes1}
	\end{center}
\end{figure*}

In designing an SRF up-conversion experiment, the
 most obvious question is what cavity geometry to use.
To detect low-frequency axions, we need a pair
of almost degenerate modes to act as the drive
and signal modes, and these should have large $C_{01}$.
If we want to use low-lying cavity modes, this generally
requires a tuning of the cavity shape.
For example, if we take a cylindrical cavity,
and consider e.g.\ the ${\rm TE}_{011}$ mode, the mode 
this is naturally degenerate with is ${\rm TM}_{111}$ (i.e.\
they are degenerate at all height-to-radius ratios).
However, due to the $m$ mismatch, these modes have $C_{01} = 0$. We can attain good overlap
with e.g.\ the $\TE_{011} / \TM_{020}$ mode pair, but the mode
frequencies are only degenerate
at $d/R \simeq 0.79$, where $d$ is the height of the cylinder
and $R$ is its radius~\cite{Goryachev:2018vjt}.
For more symmetrical cavities, such as a sphere, the overlaps
for degenerate modes are always zero.

If we are only concerned with thermal noise,
the best possible sensitivity is determined by the signal
power, and the bandwidths of the drive and signal
modes (see section~\ref{secthermal}).
Since
the signal power scales as $B_0^2$, where $B_0$ is the magnetic field
strength in the driven mode, the simplest way to increase
the signal power is to increase the amplitude of the driven mode.
However, the power dissipated increases $\propto B_0^2$
as well, and even if this is not a problem (or other issues such as field
emission), at some point the maximum
magnetic field at the cavity walls will increase past
$H_{sh} \simeq 0.2 \Tesla$. This means that, for a given
cavity
geometry, there is a maximum achievable signal power.

Given some constraints on available volume and cooling power,
we can ask how large a signal power can be obtained by optimising
the cavity geometry. 
The volume constraint is necessary; since dissipation is a wall
effect, whereas signal scales with volume (for fixed field amplitudes),
if we increase the cavity dimensions by a factor $\alpha$,
while reducing the field amplitudes to keep the signal power
constant, the dissipated power scales $\propto 1/\alpha$.
Hence, by scaling up the cavity, we can always reduce the required
cooling power for a given signal level.

Taking a cylindrical cavity geometry as our example, 
which height-to-radius ratio gives
rise to a degenerate mode pair with the best (axion-mass-averaged) signal
power per volume? If we hold $\max_{\partial V} B^2$ constant
(where $V$ denotes the volume of the cavity,
and $\partial V$ its boundary walls), then
among the low-lying modes, the best
choice is to drive ${\rm TE}_{012}$, and pick up in ${\rm TM}_{013}$,
which are degenerate at $d/R \simeq 2.35$. 
This mode pair has $C_{01} = 0.19$, and maximises
$U_0 C_{01} / V$ (for example,
driving $\TM_{013}$ and picking up in $\TM_{012}$ 
gives the same $C_{01}$, but the stored energy
in the TM mode is smaller, if we restrict the magnetic
field at the walls to be $< 0.2 \Tesla$).
The field profiles
for these modes are illustrated in Figure~\ref{figCylModes1}.
In addition to giving high signal power, this mode
pair also has attractive noise-rejection properties,
as discussed in Section~\ref{secbg}. Consequently,
we will use it as our nominal experimental setup for most of this paper.
Some features of this mode are summarised in 
 Table~\ref{tab1}, where its properties
for $R = 20\cm$ and $R = 50 \cm$ are given (at the degenerate $d/R$ ratio).
Properties for other sizes can be obtained by scaling.

\subsubsection{General constraints}

We can also consider more general cavity geometries.
Both the cooling power and $H_{sh}$ limitations are based on
the magnetic fields of the drive mode at the cavity walls.
It is not immediately obvious that these can't be made small
by a clever choice of cavity geometry. For example,
the wall \emph{electric} fields for the cylindrical ${\rm TE}_{012}$ mode
are everywhere zero (with the consequence
that field emission can be highly suppressed; see section~\ref{secfreecharge}).
However, as shown in appendix~\ref{apwall},
the wall fields can be related to the energy stored in
the cavity via
\begin{equation}
	U = \frac{1}{2}\left\langle \oint_{\partial V} dA \, (B^2 - E^2) (x \cdot n)
	\right\rangle
	\label{eqube}
\end{equation}
where $x$ is the vector from some origin to the wall location,
$n$ is the outward-pointing normal to the wall,
and the angle brackets denote time averaging.
Since $\oint dA \, x \cdot n = 3V$, if we can choose
an origin for which $x \cdot n \ge 0$ everywhere, then
\begin{equation}
	U \le \frac{3}{2} V \max_{\partial V} \langle B^2 - E^2 \rangle
	\le \frac{3}{4} V  \max_{\partial V} B^2 
	\label{equa1}
\end{equation}
for a harmonic oscillation,
so the magnetic field energy inside the cavity can be bounded
by the maximum magnetic field at the walls. Similarly,
the power dissipated is 
\begin{equation}
	P_{\rm diss} = R_s  \left\langle \oint dA \, B^2 \right\rangle
\end{equation}
Since, if $x \cdot n \ge 0$ everywhere,
\begin{equation}
	U \le \frac{1}{2} \left( \max x \cdot n \right)
\left\langle \oint dA \, B^2 \right\rangle
\end{equation}
we have
\begin{equation}
	P_{\rm diss} \ge \frac{2 R_s U}{\max x \cdot n}
\end{equation}
From equation~\ref{eqmapow}, the signal power is bounded by
the energy in the drive mode.  Consequently, 
for `simple' cavities, for which there is an interior point
from which all of the walls are visible, a given signal power
implies a lower bound on $\max_{\partial V} B^2$, and on $P_{\rm diss}$
(in the latter case, assuming a given linear extent for the experiment,
as per the scaling discussion above).

Comparing these limits to the properties of the $\TE_{012} / \TM_{013}$
mode pair, the maximum signal power per volume, for
a given $\max_{\partial V} B^2$, is $\lesssim 10$ times larger,
using the above limits. Similarly, for a cavity with the same
maximum linear extent, the signal power is $\lesssim 12$ times
larger, for a given dissipated power. 
For realistic geometries, the limits are probably
significantly lower. 

\subsubsection{Non-convex geometries}
\label{sectoroid}

\begin{figure}
	\begin{center}
		\includegraphics[width=.9\columnwidth]{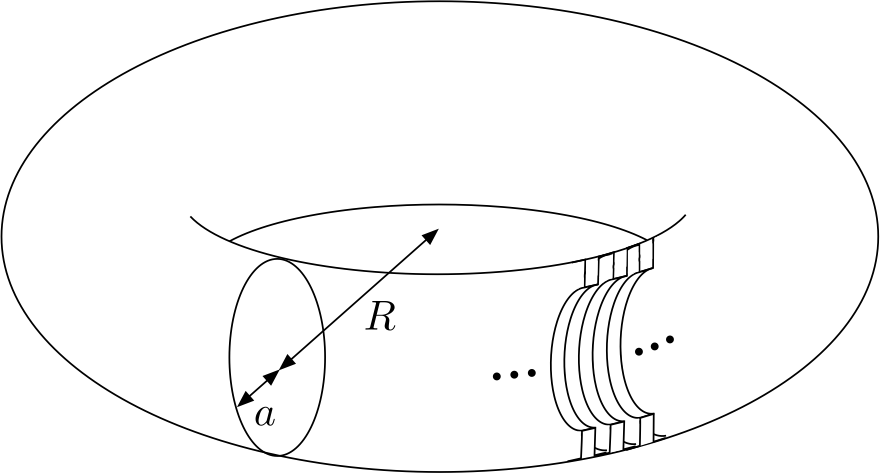}
		\caption{Illustration of a toroidal corrugated waveguide,
		as discussed in section~\ref{sectoroid}. The radius of curvature
		$R$ is taken to be much larger than the waveguide radius $a$.}
	\label{figtoroid1}
	\end{center}
\end{figure}

However, equation~\ref{eqube} also makes it fairly
obvious how to get around these limitations. If
$\oint dA |x \cdot n| \gg 3 V$, then $U$ can be
large even if the wall fields are small.
As an explicit example, we can consider a toroidal cavity 
(illustrated in figure~\ref{figtoroid1}), formed by bending a 
corrugated cylindrical waveguide of radius $a$ around to meet itself,
resulting in a toroid with overall radius $R$.
If we assume quarter-wavelength-deep corrugations, then
at frequencies for which $\omega a \gg 1$, 
the modes of a corrugated waveguide are dominantly transverse,
and the wall fields are suppressed by $\sim (\omega a)^{-1}$
relative to the fields in the interior~\cite{Doane,Clarricoats_1984,Kowalski_2010}.
Taking an explicit example, 
the linearly-polarised ${\rm HE}_{11}$ modes
have transverse fields $|E_\perp|,|B_\perp| \sim J_0 (k_0 r)$,
where $k_0 a$ is the first zero of $J_0$, and have
\begin{equation}
	\left|\frac{B(r=a)}{B(r=0)}\right| \simeq \frac{1.25}{\omega a}
	\simeq \frac{0.2}{a /\lambda}
\end{equation}
where $\lambda$ is the free-space wavelength of the mode
(and similarly for the electric fields)~\cite{Doane}.
Consequently, by making $\omega a$ large, the 
interior fields can be made parametrically larger
than the wall fields.

For a linear waveguide, the problem comes at the
end-caps --- here, the wall fields are $\sim |B(r=0)|$.
Bending the waveguide into a toroid eliminates these endcaps.
Of course, in order to preserve the smallness of the
wall fields, the radius of curvature $R$ must be large compared
to $a$. Performing a naive perturbative calculation~\cite{Miyagi_1984},
in which we take both $R/a$ and $\omega/a$ to be large,
and treating the corrugated wall as a surface with uniform effective
reactance~\cite{Doane},
the correction to the wall fields from the waveguide's
curvature is approximately
\begin{equation}
	\frac{\delta B(r=a)}{B(r=a)|_{R = \infty}}
	\simeq 0.5 \frac{a}{R} (a \omega)^2
\end{equation}
The constant factor in this estimate should be treated
as an $\OO(1)$ estimate, since we do not calculate
the actual behaviour in the corrugations.
However, the parametric form should hold, and illustrates
that
as long as $R \gg a$, it is possible
to make $a \omega$ large and obtain interior
fields significantly larger than the wall fields.

Taking some illustrative numbers, if we assume
a toroid with waveguide radius $a = 10 \cm$, 
and take $R = 4 \metre$, then choosing as high
as mode frequency as practical, $\omega \simeq 2 \pi \times 3 \GHz$, gives $a \omega \simeq 6$. 
The total energy stored
in the ${\rm HE}_{11}$ mode, for wall fields $\lesssim 0.2 \Tesla$,
is $\simeq 40 \kJ$. For a $\TE_{012} / \TM_{013}$ cylinder
of comparable volume, $U \simeq 10 \kJ$.

To use the toroidal cavity for axion detection,
we need almost degenerate modes with good overlap.
Conceptually, this is very simple --- we simply use the
${\rm HE}_{11}$ modes with orthogonal polarizations,
offset by a quarter-wavelength along the toroid. This
is in exact analogy to the optical up-conversion experiments
proposed in~\cite{DeRocco:2018jwe,Obata:2018vvr,Liu:2018icu}, which convert linearly-polarized optical photons
to the orthogonal polarization. It has the advantage of
having, in the large-$R$ limit, perfect overlap between the 
drive and signal modes. Consequently, the
advantage in axion signal strength is even larger
than the advantage in stored energy --- using the same
nominal parameters as above, the toroid's signal strength
would be $\sim 20$ times higher (a more detailed
calculation would be required to find
the proper finite-$R$ behaviour). This compares to the 
factor 10 limit derived above for `simple' cavities,
showing that the non-trivial geometry is necessary for such
improvements. If we instead used a linear corrugated
waveguide, then the drive amplitude would be constrained
by the end-caps, and the signal strength would be
$\sim$ the same as a $\TE_{012}/\TM_{013}$ cylinder
of the same volume.

The toroidal cavity serves as a particularly symmetrical,
and thus easy-to-analyse, example of a
cavity with large $\oint dA |x \cdot n|$. There are, of course,
many other possibilities. For example, we could instead 
take a linear corrugated waveguide, and replace its endcaps by
large-area reflectors. In~\cite{Bowden:2011zb}, it is claimed that this
can reduce the peak magnetic field at the walls by a factor
$\sim 2$; for a long waveguide, this would result in a
signal power per volume $\sim 4$ times larger than for
the $\TE_{012} / \TM_{013}$ cylinder pair. It is not immediately clear whether the enhancement can be made parametrically 
large, as for the toroid example.

These kinds of cavities may be significantly more complicated
to fabricate than the simple cylindrical cavities
discussed above. In addition, as we will discuss in the
next section, they lack some of the noise-rejection
properties of more symmetrical cavities, and would
be more complicated to tune and control. Consequently,
we will not attempt to analyse them in detail. 
They do, however, serve as an example of how larger signal
powers could potentially be realised.

In table~\ref{tab1}, we list some estimated parameters
for a $R = 8 {\rm \, m}$, $a = 13 {\rm \, cm}$ toroidal cavity used
as an up-conversion experiment. This size is chosen
make significant QCD axion sensitivity possible,
at least theoretically
(see section~\ref{secthermal} and figure~\ref{figComp1}).
Even these large sizes are significantly
smaller than the axion coherence
length for the masses of interest,
$l_a \sim 10^3 \nu_a^{-1} \sim 300 {\rm \, m} \frac{\GHz}{\nu_a}$, so our approximation of the axion field as spatially constant
will be valid.


\section{Backgrounds \& Sensitivity}
\label{secbg}

\subsection{Thermal and amplifier noise}
\label{secthermal}

The signal of axion DM in our experiment
would be excess power in the signal mode (as illustrated
in Figure~\ref{figupconv}),
and any noise present at these frequencies will
make this harder to detect.
As mentioned in section~\ref{secsrf}, the power dissipated
at high drive fields means that cooling the cavity to
sub-kelvin temperatures is impractical. Since $2 \pi \GHz
\simeq 50 {\rm \, mK}$, this means that the physical
temperature is always significantly higher than the signal
mode frequency, and thermal noise needs to be taken into
account. In addition, the system we use to
read out the signal --- generally, a chain of microwave
amplifiers --- will introduce its own noise.

In appendix~\ref{apsnr}, we review the theory of
signal detection for a high-$Q$ target mode,
assuming readout via an amplifier isolated
behind a circulator
(as for most microwave systems).
If the noise associated with the amplifier
system corresponds to a smaller temperature
than the physical temperature of the cavity,
then it is favourable to `overcouple' the signal mode
to the output port (i.e.\ have it lose more power to
the output port than to environmental
dissipation) \cite{sclong}.
This reduces the loaded quality factor of the target mode,
which is naively bad,
but also dilutes the thermal noise reaching the amplifier;
the latter effect turns out to dominate.

The high quality factors attainable in SRF cavities
mean that, even given this overcoupling, the bandwidth
of the target mode will be $\ll \nu_a$, for 
axion masses of interest.
 This means that,
to cover an $\OO(1)$ range in axion masses, we need
to operate the experiment in multiple different configurations,
with different frequency splittings between the drive
and target modes (as discussed briefly in section~\ref{secsrf}).
As demonstrated in Appendix~\ref{apsnr},
in the limit of long integration times, any sufficiently dense,
and roughly equal, 
spacing of these frequency splittings will give approximately the
same expected SNR,
\begin{equation}
	{\rm SNR} \simeq \sqrt{ 0.5 \frac{(P_1/Q_1)^2 t_{\rm tot} Q_a Q_1 \omega_1}{T_0 T_n m_a \Delta m_a} }
	\label{eqsnr1}
\end{equation}
Here, $T_0$ is the physical temperature of the cavity walls
(assumed to be $\gg \omega_1$),
$T_n$ is related to the noise of the amplifier system (see
appendix~\ref{apsnr}),
$\Delta m_a$ is the range of axion masses we want to cover (assumed
to be $\lesssim \OO(1)$), $Q_a \simeq 10^6$ is the inverse
fractional bandwidth of the axion
signal, $t_{\rm tot}$ is the total integration time for all configurations, $Q_1$ is the unloaded quality factor of the target mode,
$\omega_1$ is the frequency of the target mode (assumed
to change by a small fractional amount between configurations),
and $P_1$ is the power absorbed by the (unloaded) target mode when
on-resonance with a monochromatic signal (see equation~\ref{eqp1}).
Since $P_1 \propto g^2$, the smallest coupling we have sensitivity
to scales as $g_{\rm sens} \propto t_{\rm tot}^{-1/4} Q_1^{-1/4} T_0^{1/4} T_n^{1/4}$, as expected~\cite{sclong}.

From~\cite{rnl}, we know that for fixed $t_{\rm tot}$, this
improvement with increasing $Q_1$ must saturate at some point.
Fairly obviously, this will happen when the time spent in each
configuration is not long enough to resolve the loaded
bandwidth of the target mode. If, as is the case in most
of our parameter space, $\Delta \omega_a \simeq m_a/Q_a$
is such that $\Delta \omega_a \gg \omega_1 / Q_l$, where $Q_l$ is the loaded quality factor
of the target mode, then the minimum spacing of
frequency splittings is $\sim \Delta \omega_a$ (otherwise
some axion masses will fall into the gaps). This means that
the time spent in each configuration is $t_1 \simeq t_{\rm tot}
\frac{\Delta \omega_a}{\Delta m_a} \simeq \frac{t_{\rm tot}}{Q_a} \frac{m_a}{\Delta m_a}$. If $t_1 \lesssim Q_l/\omega_1$, 
then the mode cannot fully ring up in the time available,
and it would be more favourable to reduce $Q_l$ by overcoupling
further. In this regime, the best attainable SNR is
\begin{equation}
	{\rm SNR} \lesssim 0.2 \frac{\overline{W}}{T_n}
	\label{eqsnrn}
\end{equation}
where $\overline{W} \equiv \bar P t_{\rm tot}$ is the expected
power absorbed from the axion signal. 

\begin{figure*}[t]
	\begin{center}
		\includegraphics[width=.8\linewidth]{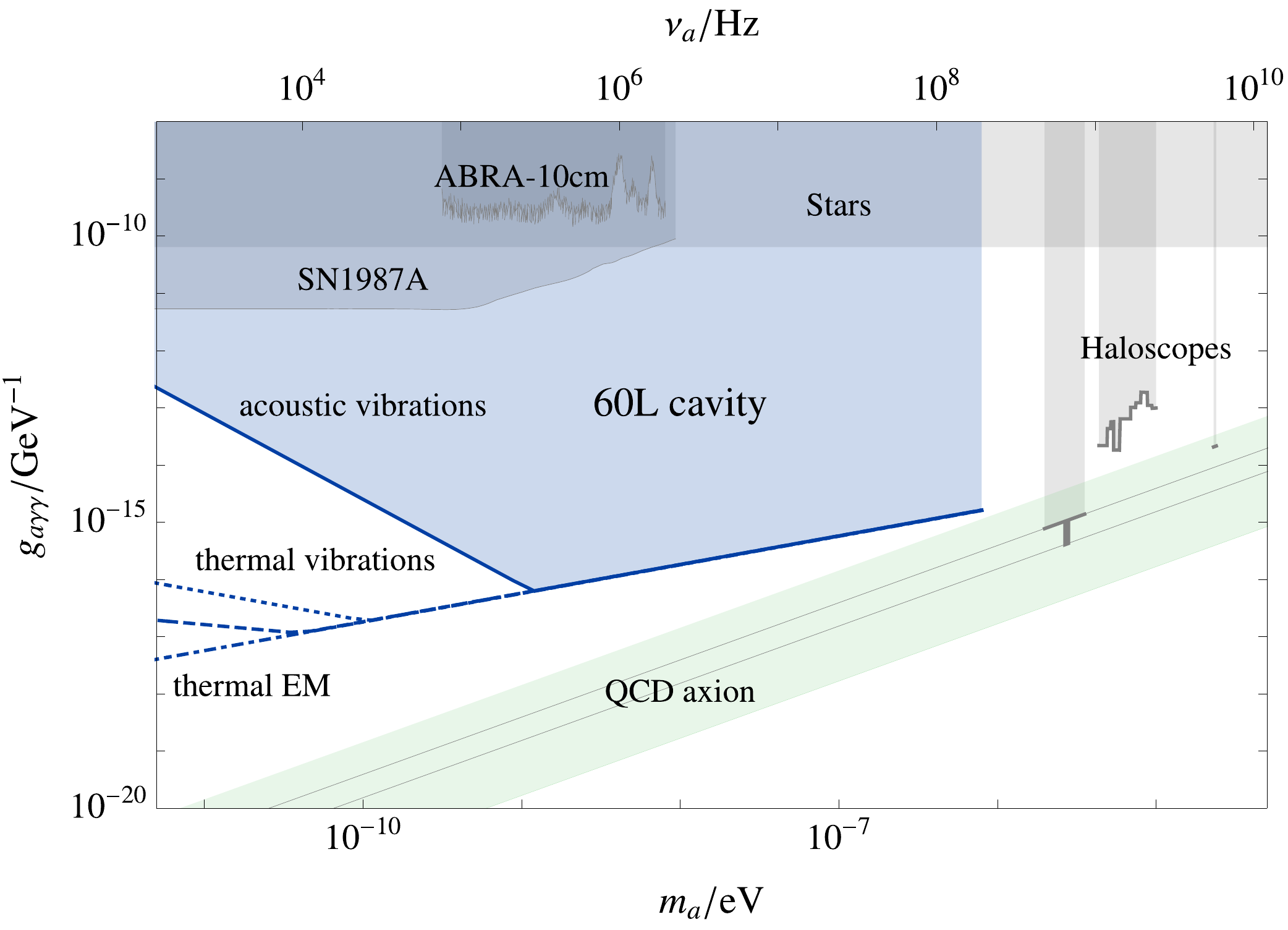}
		\caption{Sensitivity projection for an up-conversion experiment
		using a 60 litre ($20 \cm$ radius) cylindrical cavity, where
		the ${\rm TE}_{012}$ mode is driven,
		and signals are picked up in the ${\rm TM}_{013}$ mode.
		We assume that the maximum magnetic field
		for the drive mode at the cavity wall
		is $H_{sh} = 0.2 \Tesla$, and an integration
		time of one year per e-fold in axion mass range.
		The sensitivity threshold is set at an expected
		SNR value of 3.
		The `thermal EM' line shows the sensitivity limit
		in the presence of thermal EM noise,
		at the assumed physical temperature of $1.4 \kelvin$
		(see section~\ref{secthermal}).
		The `thermal vibrations' line is an estimate
		of the effect of thermal vibrations of the cavity
		walls, which up-convert power from the
		signal mode to the drive mode (section~\ref{secvibs}).
		The `acoustic vibrations' line also incorporates
		extra acoustic noise from external sources.
		The dashed line between them is an estimate
		of the effects of time-varying drive signal
		leakage (section~\ref{secleakage}).
		Taking all of these noise sources into account, the
		estimated sensitivity reach of the experiment is given 
		by the blue shaded region.
		It should be noted that, while the `thermal EM'
		limit is set by basic physical parameters, the
		vibrational limits depend on guesses
		about the properties of the cavity system, and
		could be very different in a real experiment.
		Also, while we have extended the projected
		reach to axion frequencies $\sim 100 \MHz$,
		scanning an order one axion mass range
		at these frequencies would be difficult (see
		section~\ref{secsrf}).
		In Figure~\ref{figComp1}, this reach is compared to
		other proposed experiments.
		The gray shaded regions correspond to the parameter
		space ruled out by observations of horizontal
		branch stars~\cite{Vinyoles:2015aba,Schlattl:1998fz,Ayala:2014pea}, SN1987A~\cite{Payez:2014xsa},
		existing cavity haloscope
		experiments~\cite{Brubaker:2016ktl,Zhong:2018rsr,Wuensch:1989sa,Hagmann:1990tj,Asztalos:2009yp,Du:2018uak}, 
		and the ABRACADABRA-10cm experiment~\cite{Ouellet:2018beu}.
		The green diagonal band corresponds to the `natural' range of
		$g_{a\gamma\gamma}$ values at each QCD axion mass ---
		if we write $g_{a \gamma \gamma} = \frac{\alpha_{\rm EM}}{2 \pi f_a}\left(\frac{E}{N} - 1.92\right)$~\cite{diCortona:2015ldu},
		then the upper edge of the band is at $E/N = 5$~\cite{DiLuzio:2016sbl},
		and the lower edge at $E/N = 2$~\cite{diCortona:2015ldu}.
		The gray diagonal lines indicate the KSVZ (upper, $E/N = 0$)
		and DFSZ (lower, $E/N = 8/3$) models.}
	\label{figCyl1}
	\end{center}
\end{figure*}

In figure~\ref{figCyl1}, the dot-dashed line shows
the thermal-noise-limited sensitivity for the nominal
60L cavity discussed above,
given a total integration time of one year to cover
an $e$-fold in axion mass range. It assumes that the physical temperature
of the cavity is $T_0 = 1.4\kelvin$, and the amplifier
system is quantum-limited, so $T_n = \omega_1$.
Near-quantum-limited amplifiers have been
demonstrated at microwave frequencies,
and incorporated in the ADMX~\cite{Du:2018uak,Braine:2019fqb} and HAYSTAC~\cite{Droster:2019fur} axion DM experiments.
The loaded quality factor, for the optimal overcoupling
level, is $Q_l \simeq 2 \times 10^9$, so it takes $\sim 2 \second$
to resolve the signal mode bandwidth, and
a total integration time of $\gtrsim 2 \times 10^6 \second \simeq$ 
24 days to be in the regime of equation~\ref{eqsnr1}.
Since the coupling sensitivity scales like $g_{\rm sens} \propto m_a^{1/2}$, but $g \propto m_a$ for the QCD axion, lower-mass
QCD axions are still harder to detect.
As the figure shows, an experiment with these parameters
could not attain QCD axion sensitivity over an $\OO(1)$
axion mass range, even at higher masses (without violating
our assumptions, e.g.\ by injecting a highly non-classical state
into the target mode). Moreover, though we extend the sensitivity
projection up to $\nu_a \sim 100 \MHz$, tuning a cavity
over an $\OO(1)$ range of frequency splittings would be difficult
at such high masses, as discussed in section~\ref{secsrf}.

\begin{figure*}[t]
	\begin{center}
		\includegraphics[width=.8\linewidth]{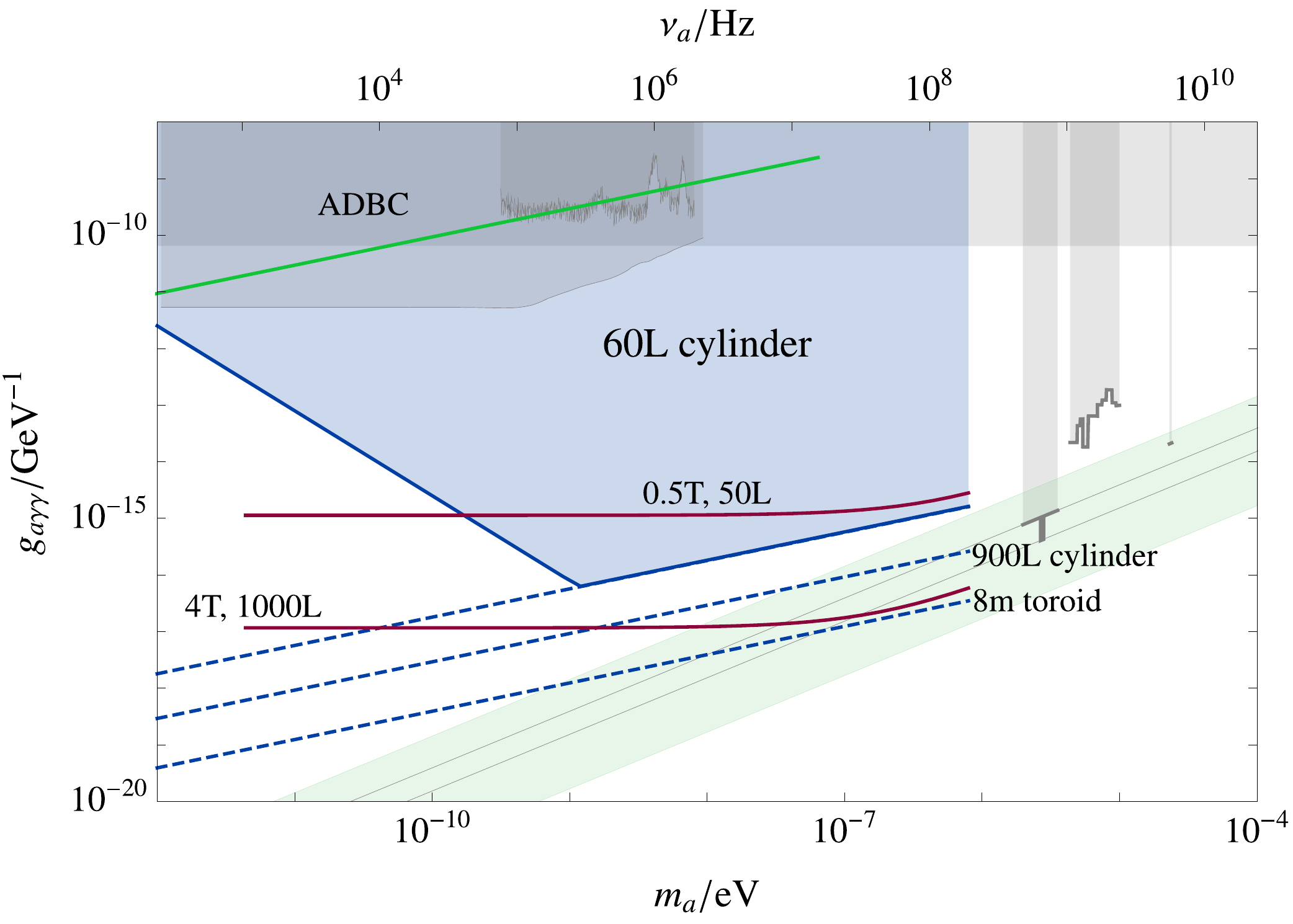}
		\caption{Comparison of sensitivity projections
		for different kinds of low-frequency axion DM detection experiments.
		The `60L cylinder' region corresponds to the
		$20 \cm$ cavity up-conversion experiment described in
		the text, with parameters given in table~\ref{tab1}
		(we assume an integration time of 1 year per
		e-fold in axion mass range).
		At high frequencies ($\nu_a \gtrsim 200 \kHz$),
		its sensitivity is limited by thermal EM
		noise from the cavity walls, while at lower
		frequencies it is limited by (a rough estimate
		of) up-converted acoustic noise from external sources (see
		section~\ref{secvibs} and figure~\ref{figCyl1}).
		We also display the thermal-noise-limited sensitivities
		for
		the larger up-conversion experiments
		listed in table~\ref{tab1}.
		The `900L cylinder' line
		corresponds to a scaled-up ${\rm TE}_{012} \rightarrow 
		{\rm TM}_{013}$ experiment, with $50 \cm$ radius.
		The `8m toroid' line corresponds to a corrugated
		toroidal cavity, as described in section~\ref{sectoroid};
		we assume an overall radius of $8 \metre$,
		a waveguide radius of $13 \cm$, and a frequency
		of $2.8 \GHz$, resulting in the parameters listed
		in table~\ref{tab1}.
		We leave an analysis of the less fundamental noise sources
		for these larger up-conversion experiments to future work.
		The red lines are thermal noise limited sensitivity
		projections for representative static-magnetic-field
		experiments, modelled on the Dark Matter
		Radio proposal~\cite{scshort,dmr}.
		The `0.5T, 50L' line corresponds
		to a 50 litre experiment, with a background
		magnetic field of $0.5 \Tesla$, and a 
		physical temperature of $10 {\rm \, mK}$
		--- the resonator quality factor is taken to be
		$10^6$, and the geometric overlap
		factor to be 0.2~\cite{scshort} (in the sense
		that $P_{\rm sig} \simeq 0.2^2 g^2 B_0^2 V Q \frac{\rho}{m} (m L)^2$, where $L$ is the shielding length scale~\cite{rnl}). The `4T, 1000L' line
		corresponds to the same assumptions, but with
		a cubic metre volume, and a background
		magnetic field of $4 \Tesla$.
		The quasi-static $\sim (m L)^2$ suppression
		used is only a dimensional estimate;
		the actual value would depend on the geometry of
		the experiment.
		We also show an example of a proposed optical-frequency
		up-conversion experiment, the `ADBC' proposal
		from~\cite{Liu:2018icu}. This assumes a 
		2 metre Faby-Perot cavity, with a circulating optical
		power of $10 {\rm \, kW}$.
		The existing haloscope limits, QCD axion band,
		and astrophysical constraints are as per Figure~\ref{figCyl1}.}
	\label{figComp1}
	\end{center}
\end{figure*}

The simplest way to obtain higher sensitivities would be to use
larger cavities. Figure~\ref{figComp1} shows the thermal-noise-limited sensitivities for two such examples. The first is a 
simple scaling-up of
the $\TE_{012}/\TM_{013}$ configuration, using a
cylinder of radius $50\cm$ (giving a volume of $\sim 900 \litre$).
Its properties are summarised in Table~\ref{tab1} ---
compared a smaller cylinder, the signal power
scales with the volume. Even an experiment of this size
could not reach KSVZ axion sensitivity over an $\OO(1)$
axion mass range, at reasonable
axion masses. At this size, the mode frequency is $\sim 440 \MHz$,
which is at the lower end of the frequency
range
generally used in SRF cavities.

To illustrate the kind of parameters that would be
required for significant DFSZ axion sensitivity,
Figure~\ref{figComp1} also shows the approximate thermally-limited
sensitivity for a corrugated
toroidal cavity, as introduced in section~\ref{sectoroid}.
This is taken to have waveguide radius $a = 13\cm$,
and $R = 8 \metre$; the resulting properties are summarised
in Table~\ref{tab1}. 
The naive estimate of the signal mode quality factor
is high enough to put us in the regime of equation~\ref{eqsnrn},
even for a total integration time of a year.

As emphasised in section~\ref{sectoroid},
this kind of projection should not be taken
as a concrete experimental proposal --- for that, one would need
to understand the control issues, noise problems etc.\
associated with these cavity designs. Instead, it illustrates
what would be necessary, in principle, to probe
very small couplings.

\subsection{Vibrations}
\label{secvibs}

While thermal noise must be taken into account in all
axion detection experiments, up-conversion
experiments have the additional challenge of a
very large-amplitude background EM oscillation.
If there are any environmental oscillations
at the axion frequency, and any non-linear processes
which couple these to the drive and signal modes,
then power can be delivered to the signal mode,
representing an additional noise source.

The most obvious such process is mechanical vibrations
of the cavity walls. If we write $\delta x(t, x)$
as the displacement of the cavity wall
from initial position $x$, and $n$ as the outward pointing
normal to the initial wall position,
then to linear order in $\delta x$, the interaction
Hamiltonian of the wall displacement with the EM fields
is 
\begin{equation}
	H_{\rm int} = \frac{1}{2} \int dA (B^2 - E^2) n \cdot \delta x
\end{equation}
The resulting interaction between the drive and signal
modes is 
\begin{equation}
	H_{\rm int} = \int dA (B_0 \cdot B_1 - E_0 \cdot E_1) n \cdot \delta x
	\label{eqhintwalls2}
\end{equation}
This interaction actually represents the signal mechanism for 
proposed SRF gravitational-wave detectors, such
as MAGO~\cite{Bernard:2001kp,Ballantini:2003nt,Ballantini:2005am} --- the effect of a gravitational wave
can be treated as a deformation of the cavity
walls, which up-converts photons from the drive mode
to the signal mode.

We can see from equation~\ref{eqhintwalls2} that,
if the drive and signal modes of the unperturbed
cavity are orthogonal at the walls, then to linear
order in $\delta x$, wall vibrations do not couple them.
This is one reason why the $\TE_{012}/\TM_{013}$ mode
pair introduced in section~\ref{seccavgeom} is attractive 
--- in an ideal cylinder, these modes have
$E_0 \cdot E_1 = 0$, $B_0 \cdot B_1 = 0$ throughout
the whole cavity.
However, it will not be possible to fabricate the cavity shape
perfectly, and some deformations will result in non-orthogonal
fields at the walls. For example, if the cavity cross-section
is elliptical rather than cylindrical, then 
the deformed $\TE_{012}$ mode has a small
electric field
at the cylinder's walls,
\begin{equation}
	E_r \simeq 40 {\rm \, kV \, m^{-1}} \frac{f}{10^{-3}}
	\sin (2\phi) \sin \left(\frac{2 \pi z}{d}\right)
	\label{eqelecf}
\end{equation}
where $f$ is the flattening of the ellipse
($f = (a-b)/a)$, where $a \ge b \ge 0$ are the axis lengths),
and the normalisation is set by taking the maximum
magnetic field at the walls to be $= 0.2 \Tesla$.
Since the $\TM_{013}$ mode has
$E_r \propto \sin (3 \pi z / d)$ at the walls,
the effect of a cavity vibration
is controlled by
\begin{equation}
\int dA 
\sin (2 \phi)
\sin \left(\frac{2 \pi z}{d}\right)
\sin \left(\frac{3 \pi z}{d}\right) n \cdot \delta
x \equiv 0.09 C A_w x(t)
\end{equation}
where $A_w$ is the area of the cavity walls,
and we have normalised so that 
a vibration with $\delta x = 
\sin (2 \phi)
\sin \left(\frac{2 \pi z}{d}\right)
\sin \left(\frac{3 \pi z}{d}\right) x(t)$
has $C = 1$. Then, for monochromatic oscillations,
the power transferred to the signal mode is
\begin{equation}
P
	\simeq 7 \Watt \, C^2 Q_1 \left(\frac{f}{10^{-3}}\right)^2 \left(\frac{x}{{\rm mm}}
	\right)^2
\end{equation}
where we have taken the parameters of
the $20\cm$ cylindrical cavity listed in table~\ref{tab1}.
Consequently, the displacement noise in
the relevant bandwidth must be very small, in order
not to overwhelm the axion signal power.

The example of an elliptical cavity illustrates the general 
feature that, for a cavity deformation
of fractional size $\sim f$, we expect the wall fields
to be perturbed by $\sim f$~\cite{Bernard:2002ci,meidlinger2009general}, and
in general to be non-orthogonal.
It may be possible to deform the cavity post-fabrication
to alleviate such issues, but we do not
attempt to consider such possibilities
here.
Conversely, for other cavity geometries, such as 
a corrugated waveguide, the wall fields
are non-orthogonal even for an unperturbed cavity.

The other information we need to determine the 
vibration-induced noise spectrum is the 
vibration spectrum of the cavity's walls.
At high enough frequencies, this will probably be dominated by thermal
vibrations. 
If ultrasonic waves from outside the cavity are 
sufficiently attenuated, then the amplitude will
be set by the cavity's physical temperature.
The ultrasonic attenuation length scale in liquid helium is $\sim$ cm 
for frequencies $\gtrsim 10 \MHz$~\cite{Whitney_1957,Jeffers_1965},
though it may be significantly smaller for metals
\cite{Whitney_1957,Tsuda_1967}.
Investigation would be required to determine the actual
properties of a cavity setup.

At lower frequencies, external acoustic noise will not be
strongly attenuated, and in addition, above-thermal
noise sources (such as machinery) will be present.
A nominal spectrum for this external displacement noise,
in a reasonably `quiet' setting, is
\begin{equation}
	\sqrt{S_{xx}(\nu)} \simeq 10^{-7} \frac{\cm}{\sqrt{\Hz}}
	\left(\frac{10 \Hz}{\nu}\right)^2 
	\label{eqsxx1}
\end{equation}
for $\nu \gtrsim 10\Hz$~\cite{Saulson:2017jlf}.
The measured acoustic noise in the vicinity of 
the MAGO prototype (figure 14 of~\cite{Ballantini:2005am}) also shows a $S_{xx} \sim 1/\nu^4$
spectral density for $\nu \gtrsim 3 \kHz$, with displacement amplitude a factor $\sim 10$ higher than equation~\ref{eqsxx1}.
At lower frequencies, especially with the helium cooling
system in operation, there is a complicated, spiky spectrum
with significantly larger amplitude.

We can compare these acoustic spectra to the spectrum
from thermal oscillations by noting that, 
if we can treat the material as a 
bulk medium of large extent, with a linear acoustic
dispersion relation, then the PSD of surface
displacements is given by~\cite{Weaver_2001}
\begin{equation}
	S_{xx} \simeq \frac{2 p T}{9 \pi 
	\rho v_s^3}
\end{equation}
where $\rho$ is the density of the material,
$T$ is the temperature, $v_s$ is the sound speed,
and $p$ is a dimensionless constant governing the interaction
of the surface with bulk phonons (e.g.\ for aluminium,
$p \simeq 2.3$~\cite{Weaver_2001}).
Plugging in representative parameters for
niobium, we obtain
\begin{equation}
	\sqrt{S^{\rm therm}_{xx}} \simeq 
	10^{-17} \frac{\cm}{\sqrt\Hz} \left(\frac{T}{1.4 \kelvin}\right)^{1/2}
\end{equation}
suggesting that, very roughly,
thermal noise will dominate for $\nu \gtrsim \MHz$.

Properly calculating the effects of vibrations would require
assumptions about the imperfections of the cavity shape, and
modelling the mechanical modes of the cavity-cryostat system.
Here, we only attempt to make a rough estimate
of these effects.
To calculate the vibration-induced noise
power as a function of frequency,
we need the PSD $S^v_{xx}$ for oscillations
with profile set by $B_0 \cdot B_1 - E_0 \cdot E_1$
at the cavity walls. We will assume
that this profile is reasonably smooth, corresponding
to slowly-varying deviations from a cylindrical
shape (e.g.\ the ellipse example above).

Decomposing the cavity vibrations into weakly-damped
modes, there will be contributions to $S^v_{xx}$
from modes with different resonant frequencies.
At $\omega$ much higher than the resonant
frequency of low-order mechanical modes,
the vibrational modes with similar frequencies
will have small overlap with the (assumed smooth)
profile, while the low-order vibrational modes
will have much smaller frequencies. For thermal
vibrations, the contribution
from the latter will be $S_{xx} \sim
\frac{1}{Q_m} \frac{T}{M \omega^3}$, where
$M$ is the mass of the cavity walls,
and $Q_m$ is the mechanical quality
factor of the low-order modes.
The modes with resonant frequency
$\sim \omega$ will contribute $S_{xx} 
\sim (\lambda / L)^c \frac{T}{M \omega^3}$, 
where $\lambda \sim v_s/\omega$ is the wavelength of
 the modes,
and $ (\lambda / L)^c$ corresponds
to the suppressed spatial overlap with the profile
(with higher $c$ corresponding to smoother profiles).
At frequencies low enough that individual modes
stop overlapping in frequency, the contributions
to $S_{xx}$ will be spikier. For $\omega$ around
that of the low-lying mechanical modes,
it will have maximum value $S_{xx} \sim Q_m \frac{T}{M \omega^3}$.

For acoustic, rather than thermal, noise,
the same general considerations will apply,
but the effective $T$ in the above equations will
be frequency dependent. 
In figure~\ref{figCyl1}, 
we display an example of the vibration-noise-limited
sensitivity, using the estimates discussed above.
Taking acoustic noise with spectrum
as per equation~\ref{eqsxx1}, we find that 
vibration-induced noise dominates thermal EM noise
for $\nu_a \lesssim 300 \kHz$ (we assume that $Q_m \simeq 10^3$ and $c = 2$,
in the notation of the previous paragraph,
and assume $3 {\rm mm}$ thick cavity walls, giving
a cavity mass of $\sim 20 {\rm \, kg}$).
The thermal vibrations, assuming $T = 1.4 \kelvin$,
are sub-dominant.
We emphasise that all of these estimates are best
viewed as guesses, and a much more careful treatment
would be necessary for a realistic experiment.

\subsubsection{Frequency variability}
\label{secfreqvar}

So far, we have considered how vibrations
can give rise to a coupling between
the drive and signal modes. However, vibrations will
also change the energy of each of these modes themselves.
Typically, this `detuning' will (before any
compensation) be $\OO(10 \Hz)$, corresponding
to $\sim$ nm-scale displacements of the cavity walls
\cite{Ge:2017nwd,Kelly:2003zz,Neumann:2010zz}.
Most of this variation will come from
low-frequency ($\lesssim \OO(10 \Hz)$) vibrations.

If these vibrations cause the frequency splitting between
the drive and signal modes to change over time, 
by more than their bandwidths, then we are effectively
changing the up-conversion tuning
over time. This change in `scan strategy' can affect the thermal
noise limited sensitivity, as discussed in section~\ref{secthermal}.

For SRF accelerator cavities, large frequency detunings
can increase the power needed to drive the cavity~\cite{Padamsee:2015asa}, as well
as moving the accelerating voltage out of phase with the
electron bunches~\cite{Neumann:2010zz}. Consequently, feedback systems are
general employed which measure the detuning of the cavity,
and apply a mechanical deformation (via a piezo transducer)
to correct this. Using such systems, frequency stabilization
to $\lesssim \Hz$ has been demonstrated~\cite{Neumann:2010zz}
In our case, measurement of the changing mode frequencies
is very important, since any \emph{unknown} variation of
the frequencies over time will reduce sensitivity.
Further investigation of how well such a measurement system
(and feedback control system) could perform 
in our setup would be required.

In principle, it may also be possible to measure
the higher-frequency mechanical vibrations of the cavity
discussed above,
and use this to compensate for, or subtract out, the induced
noise in the signal mode. We do not attempt to model the
feasibility of such measurements here.


\subsection{Drive leakage}
\label{secleakage}

In an ideal experiment, the cavity input power would
couple only to the drive mode, and the detector would
be coupled only to the signal mode. 
For mode pairs such as our $\TE_{012} / \TM_{013}$ example,
it is easy to see how this could
be accomplished in principle.
Input and output is usually accomplished through small slots
in the cavity wall, which connect to waveguides.
For a slot much smaller than the wavelength of a cavity
mode, the coupling is approximately $\propto B_{\rm slot} \cdot 
B_{\rm cav}$, where $B_{\rm cav}$ is the mode's magnetic field
at the cavity wall~\cite{Alesini:2011aa}. Since the signal and drive
modes are everywhere orthogonal, a correctly polarized input/output
port should, to a good approximation, couple only to one
of them. The different spatial profiles of the modes
(see Figure~\ref{figCylModes1}) provide an extra
discriminator --- for example, we could place an input
port at a wall location where the signal mode's
magnetic field is very small, and an output port
where the drive mode is small.

However, as was the case for the vibrational couplings discussed above,
non-perfect fabrication of the cavity will spoil
these assumptions. For example, if the output slot
was slightly misaligned, it would have a small coupling
to the drive mode. 
Generally, for a Fourier-domain signal $S_I(\omega)$
at the input port, and assuming that the drive and signal
modes are the only ones at nearby frequencies,
the output signal will be 
\begin{equation}
	S_O(\omega) \simeq C_{oD} G_D(\omega) C_{Di}
	+ C_{oS} G_S(\omega) C_{Si}) S_I(\omega)
\end{equation}
Here, $C_{oD}$ represents the coupling between the 
drive mode and the output port, and
$G_D(\omega)$ is the response function of the drive mode, etc.
For a well-constructed cavity, $C_{oD}$ and $C_{Si}$
will be small, but still non-zero.
Since the signal and drive modes have very narrow-bandwidth
response functions, we expect $S_O(\omega)$ to
be peaked around $\omega_0$ and $\omega_1$.

Naively, the output signal near $\omega_0$ is not an
issue --- we can just Fourier transform the output
data, and ignore it. However, the non-ideal nature of
the output electronics may introduce problems. Most
directly, if the amplitude of the $\sim \omega_0$
noise, relative to the $\sim \omega_1$ signal, is
too large, then the dynamic range of the amplifier
may be exceeded. One way to address this is to place
a frequency filter on the output port, to reject
frequencies too far from $\omega_1$.
However, if this is not enough (which is especially likely
at smaller $m_a$), it may be necessary to reduce
$C_{oD}$ through some feedback control mechanism.

Using the nominal experimental parameters
from above, the energy stored in the drive
mode of the 60L cavity is $\sim 700 \Joule$.
If the output port were critically coupled to the drive mode, this would result in
$\sim 30 \Watt$ output power. At the 
thermal-noise-limited sensitivity
(section~\ref{secthermal}), the axion-sourced
signal power is $P_{\rm sig} \simeq 2 \times 10^{-23} \Watt$
for $\nu_a \sim \MHz$ (assuming an integration time of 1 year per e-fold in axion mass range). If we assume an amplifier
with a dynamic range of 50dB, then to avoid swamping this signal, the output
power needs to be a factor of $\sim 10^{-20}$ lower
than the $30 \Watt$ level, corresponding to a
$\sim 10^{-10}$ amplitude suppression.
These numbers make it clear why leakage suppression might
be challenging.
Furthermore, in addition to leakage through the cavity,
we would also need to worry about leakage
through the laboratory environment. An experimental rule of thumb
is apparently~\cite{SaptarshiPersonal} that suppressions
of more than $\sim 10^{-19}$ in power are difficult
to achieve, for electronics in the same laboratory
space. 

These kinds of issues were encountered by
the MAGO SRF gravitational-wave detector mentioned
in the previous section. Their setup
used two identical cavities with a small coupling between
them, giving rise to almost degenerate symmetric and
antisymmetric modes.
To drive the symmetric mode, a magic-tee was used
to split the drive signal into in-phase components,
while to sample the antisymmetric mode,
another magic-tee was used to take the difference
of outputs from each cavity. Since
the drive and signal modes had the same profiles in each cavity,
this relative phase was the only way of distinguishing
between them.
With this setup, they achieved a 
power suppression of $\sim$ 48dB for the output,
corresponding to an amplitude suppression
of $\sim 1/250$~\cite{Bernard:2001kp}. The output power was dominated
by the $\sim \omega_0$ peak.

To improve this, they implemented a feedback system, using 
a variable phase shifter to control the phase of one of the
input ports (and another to control the phase of an output port).
This phase shifter was controlled by a feedback loop,
designed to suppress the output power. Using this system,
they achieved an output power suppression of $\sim$ 140dB
\cite{Bernard:2001kp}
(which would seem to indicate that the amplitudes
at the two output lines were matched to one part in $\sim 10^7$
by the geometry of the setup, unless they also
applied feedback control the amplitudes).
One could imagine other ways of
implementing feedback schemes --- for example,
combining the (attenuated) drive signal with the
output directly --- but the MAGO scheme provides
a concrete example of the kind of control system
that may be required. In our case, the extra
geometric rejection provided by the profiles
of the drive and signal modes, along with a
high-$Q$ bandpass filter on the output, 
make it seem plausible that sufficiently
high rejections could be obtained.

If the drive signal has narrow bandwidth compared 
to the frequency splitting between the drive
and signal modes, then leakage at frequencies
close to $\omega_1$ will be suppressed. However,
it may interfere with the axion signal we hope
to detect, so more care is required in dealing with it.
Since we can measure the input drive signal,
it does not necessarily represent irreducible noise
--- if we know the transfer function from the input
to the output, e.g.\ via the signal mode
of the cavity, then we can simply subtract it out.
This can either be done in software, or via analogue mixing (though in the former case, dynamic range issues 
may still arise). 

The more worrying case is if the transfer function
varies over time, in an a-priori
unknown way (for example, MAGO observed
that temperature fluctuations in their input
and output cables affected the phase shift experienced
by the signal~\cite{Ballantini:2005am}). As a quantitative example, suppose
that the time delay $\delta(t)$ along the signal
path is time-varying, $\delta(t) = \delta \cos \omega_\delta t$.
For $\omega_0 \delta \ll 1$, we can use
the Jacobi-Anger expansion~\cite{abramowitz_stegun},
\begin{equation}
	e^{i \omega_0 (t + \delta \cos \omega_\delta t)}
	= e^{i \omega_0 t}\left( J_0(\omega_0 \delta)
	+ 2 i J_1 (\omega_0 \delta) \cos \omega_\delta t +
	\dots \right)
\end{equation}
So, given leakage at frequencies $\sim \omega_0$,
the amplitude
of the noise component at frequencies close to $\omega_1$
is $\sim (\omega_0 \delta)$ times the constant leakage amplitude.
This gives a noise power, for our nominal setup, of
\begin{equation}
	P_{\rm noise} \sim 3 \times 10^{-21} \Watt \left(\frac{|C_{Si}|}{10^{-3}}\right)^2 \left(\frac{\MHz}{\nu_a}\right)^2 (\omega_0 \delta)^2
\end{equation}
where we take $|C_{Si}|$ to be the amplitude suppression,
compared to critical coupling, of the output ports to the
drive mode.
Translating this into an equivalent noise temperature,
\begin{equation}
	T_{\rm eff} \simeq  1 \kelvin \left(\frac{|C_{Si}|}{10^{-3}}\right)^2 \left(\frac{50 \kHz}{\nu_a}\right)^3 Q_\delta (\omega_0 \delta)^2
\end{equation}
for a $\delta$ PSD of fractional bandwidth $1/Q_\delta$,
and total power $\delta^2$.
Consequently, unless the temporal variation in
transfer characteristics is substantial,
this noise contribution is likely to be smaller
than e.g.\ that of the acoustic noise considered
in the previous section.
In figure~\ref{figCyl1}, we show the effect
on sensitivity for $Q_\delta \simeq 1$, $\omega_0 \delta \simeq 0.1$,
illustrating that, for these parameters, it is fairly similar
to our estimate for thermal vibrations.

Of course, as well as this in-principle issue, there
may be technical issues in suppressing the $\sim \omega_1$
leakage, especially for low-dynamic-range amplifiers.
For the purposes of our sensitivity estimates, we
will assume that these are surmountable at high axion masses,
and sub-dominant to vibrational noise at lower ones.

If there are other modes close-to-degenerate
with the drive and signal modes, then these may
affect signal leakage. For example, in a perfect
cylindrical cavity, the $\TE_{012}$
mode is exactly degenerate with the
$\TM_{112}$ mode. While such degeneracies
will be lifted by the inevitable cavity shape
imperfections (e.g.\ the $\sim 10^{-3}$ fractional
deformations we were assuming would generically
give $\sim \MHz$ splittings), if they are still a problem, 
then it may be simplest to give the cavity an intentional,
slight deformation, as was done for
the cylindrical MAGO prototype~\cite{Bernard:2001kp}. 

\subsection{Free charges}
\label{secfreecharge}

As well as wall effects, such as the vibrational backgrounds
considered in Section~\ref{secvibs}, EM noise can also arise through
currents inside the volume, i.e.\ from the movement of free charges
inside the cavity. These charges may originate from the cavity
walls via field emission, or may come from outside the cavity.

For initially low-energy charged particles, such as those
arising from field emission, the oscillating drive field inside the
cavity will have a significant effect on their motion, and it is necessary to solve for their trajectories inside the cavity.
However,
for particles with sufficiently high initial kinetic energy,
the effect of the drive field on the trajectory is unimportant,
and the particle effectively travels in straight line.
This is the case for e.g.\ cosmic ray muons, which 
provide a simple test case we can work out.

We are interested in the effect of the charged particle
on the signal mode. Working in a gauge where $A_0 = 0$, 
we can write the interaction Hamiltonian as 
\begin{equation}
	H_{\rm int} = \int dV \, A \cdot J
	= q \, v(t) \cdot \hat A(x(t))
\end{equation}
If we write the vector potential for the signal
mode as $\vec A(t,x) = A_s(t) \vec a(x)$, then
the dynamics of $A_s$ are analogous to those of a harmonic
oscillator with `mass' $= \int dV a^2 \equiv V_a$~\cite{rnl}.
Writing
$H_{\rm int} = (q v(t)\cdot a(x(t))) A_s(t) \equiv j(t) A_s(t)$,
the linear response function governing the response of $A_s$ to
the forcing $j$ is 
$\tilde \chi (\omega) \simeq \frac{V_a^{-1}}{\omega_1^2 - \omega^2 + i \omega \omega_1 / Q_1}$. The expected energy delivered to
the signal mode
is 
\begin{equation}
	\langle W \rangle = \frac{1}{2\pi} \int_{-\infty}^\infty d\omega \, \omega |\tilde j(\omega)|^2 {\rm Im} \tilde\chi (\omega)
\end{equation}
where the averaging is taken over different phases of the 
signal mode oscillation.
For a high quality factor mode, ${\rm Im} \tilde \chi$ will be
very narrowly peaked compared to $\tilde j$, so
\begin{equation}
	\langle W \rangle \simeq \frac{1}{2 V_a} |\tilde j(\omega_1)|^2
\end{equation}
Consequently, for a low-lying mode of the cavity, a single
relativistic transit of a charged particle will deposit,
on average, $\sim C q^2$ photons into the signal mode,
where $q$ is the charge of the particle, and $C$ is a
dimensionless geometric overlap factor. It should be noted
that, in the presence of an oscillation of definite phase in
the signal mode (for example, due to thermal fluctuations),
there will also be an $\OO(q)$ contribution to the energy
absorbed. However, the energy
uncertainty of a coherent state is also larger, and the
relative detectability of the perturbation is still set by
$\langle W \rangle/T$.

The flux of cosmic ray muons at sea level is $\sim 1/((10 \cm)^2 \second)$~\cite{pdg}. They have an average energy of $\sim 4 \GeV$, so are certainly high-energy for the purposes of our calculation. Considering
our nominal $20 \cm$-radius cylindrical cavity,
there are $\lesssim 20$ muons passing through the cavity 
per second, corresponding to a delivered power
of $\sim \bar C \times 10^{-24} \Watt$ to the signal 
mode, where $\bar C$ is the average value of $C$
over different trajectories. 
A large value of $C$, e.g.\ for a 
trajectory along the central z-axis, is $C \simeq 0.17$.
In comparison, the thermal-noise-limited
signal power, assuming an integration time 
per e-fold in axion mass range of one year,
is $P_{\rm sig} \simeq 2 \times 10^{-23} \Watt \frac{\nu_a}{\MHz}$.
Since we do not expect the flux of other high-energy charged
 particles to be significantly larger than the muon flux,
we can conclude that cosmic rays will not be
a significant background at higher $\nu_a$.
These calculations would apply to 
standard cavity haloscopes as well, with similar results
(see also~\cite{caspers}).

For cavities with small enough electric fields at the walls,
field emission should be negligible. 
Taking the previous example of a slightly elliptical
cavity, if the $\TE_{012}$ drive mode has maximum magnetic field
of $0.2 \Tesla$ at the walls, then from equation~\ref{eqelecf},
the maximum wall
electric field is $E_{\rm max} \simeq 40 {\rm \, kV/m} \frac{f}{10^{-3}}$, where $f$ is the flattening of the ellipse.
Since electric fields $\gtrsim {\rm few} {\rm \, MV/m}$ are required for
field emission, even fairly loose mechanical tolerances
should be enough to suppress it.
This is contrast to the modes used in SRF
cavities for particle acceleration, such as the
$\TM_{010}$ mode --- to obtain an accelerating
voltage down the axis of the cavity, these
have large electric fields at the walls.

Other cavity geometries, such as a corrugated waveguide, 
do not necessarily have such suitable drive
modes with small electric fields at the walls.
In these cases,
proper simulations would need to be carried out to
determine whether the resulting electron trajectories
transfer too much energy to the signal mode.
If the rate of field emission is high enough,
then multiple electrons could contribute coherently to
such energy transfer, potentially worsening the problem.


\section{Sensitivity comparisons}

\begin{table*}
	\begin{center}
		\begin{tabular}{c|cccccccc}
			& $V$ & $f$ & $U_{\rm max}$ & $C_{01} U_{\rm max}$ &
			$P_{\rm max}$ & $Q_0$ & $Q_1$ & $B_{\rm rms}$ \\ \hline
			`60L cylinder' & $60 {\rm \, L}$ & $1.1 \GHz$ & $690 \Joule$
			& $132 \Joule$ & $27 \Watt$ & $1.8 \times 10^{11}$ &
			$9.5 \times 10^{10}$ & $0.12 \Tesla$ \\
			`900L cylinder' & $920 {\rm \, L}$ & $440 \MHz$ & $11 {\rm \, kJ}$
			& $2 {\rm \, kJ}$ & $170 \Watt$ & $1.8 \times 10^{11}$ &
			$9.5 \times 10^{10}$ & $0.12 \Tesla$ \\
			`8m toroid' & $\sim 2700 {\rm \, L}$ & $2.8 \GHz$ & $\sim 210 {\rm \, kJ}$
			& $\sim 210 {\rm \, kJ}$ & $\sim 650 \Watt$ & $\sim 6 \times 10^{12}$ &
			$\sim 6 \times 10^{12}$ & $\sim 0.3 \Tesla$ \\
		\end{tabular}
	\caption{Parameters for the nominal SRF cavity experiments referred
		to in Figure~\ref{figComp1}. In each case, the maximum magnetic field at the cavity walls is taken to be $0.2 \Tesla$,
		and the surface resistance of the cavity
		walls is taken to be $R_s = 5 {\rm \, n\Omega}$. The various quantities are defined in section~\ref{secupconv}.
		Note that the quality factors given are in the sense
		of dissipation, i.e.\ $P_{\rm diss} = \omega U / Q$,
		rather than frequency stability. The latter
		will depend on how well vibrations can be
		controlled, as per section~\ref{secfreqvar}.
		The $B_{\rm rms}$ column shows the RMS
		magnetic field, averaged over time and space.
		}
	\end{center}
	\label{tab1}
\end{table*}

As mentioned in the introduction, there are a number
of different approaches to low-frequency axion
DM detection through the $F \tilde{F}$ coupling.
Whether SRF up-conversion experiments are worth pursuing
depends on their plausible sensitivity, relative to these alternatives.

Static background field experiments, such as 
the ABRACADABRA~\cite{Kahn:2016aff} and DM Radio~\cite{scshort} proposals,
offer the best theoretical sensitivity at higher
axion masses. In Figure~\ref{figComp1}, we show
thermal-noise-limited sensitivity projections
for nominal static field experiments, 
modelled on the DM Radio proposals\footnote{
	As discussed in \cite{rnl}, the quantum-limited
	sensitivity
	of ABRACADABRA-style experiments may theoretically
	be better, but it is less clear whether it
	is plausibly reachable. The DM Radio projections
	are more closely related to the SQL-limited
	sensitivity for a static-field experiment,
	so provide a simpler point of comparison here.
} \cite{dmr}, and compare
them to the sensitivity projections for the nominal
SRF experiments discussed in the previous section, with
parameters summarised in Table~\ref{tab1}.
Compared to the nominal static field parameters, the 
corresponding SRF experiments have significantly
smaller RMS magnetic fields, but gain by avoiding the 
quasi-static geometric suppression, and by having higher
target mode quality factor. Correspondingly, they have
different scaling of sensitivity with axion mass.

Figure~\ref{figComp1} illustrates that, for significant QCD axion 
sensitivity in a SRF up-conversion experiment, either very large (many cubic meters)
conventional cavities, or few cubic meter advanced
cavities, would be required (ignoring potential
quantum enhancements).
This is true even if we only
take thermal noise into account --- further study would be
required to determine whether other noise sources
could be mitigated. As discussed in the previous section,
such mitigations will likely work best at higher axion 
masses; more realistic sensitivity projections for the larger
cavities would likely have shapes similar to the small-scale cylinder
projection of figure~\ref{figCyl1}.

The projections of~\cite{slac}, which proposes
similar SRF up-conversion experiments, are presented
somewhat more optimistically, particularly
regarding QCD axion sensitivity. This seems
to be mostly due to two factors. The first is that
they take their sensitivity threshold as
${\rm SNR} = 1$, rather than ${\rm SNR} = 3$ as
we use. The second is that they present their
projections for a nominal cavity with peak spatial-RMS
magnetic field of $0.2 \Tesla$, and an overlap
factor $C_{01} = 1$. This gives a signal power
per volume $\sim 6$
times better than our $\TE_{012}/\TM_{013}$ cylindrical mode
pair (which has $\sim$ the same signal
power as the corrugated cylindrical cavities
discussed in~\cite{slac}). As we discussed in section~\ref{seccavgeom},
such a high signal power likely requires more complicated,
non-convex cavity geometries,
such as our corrugated toroids.

We can also compare SRF up-conversion experiments
to those at optical frequencies, as proposed in
\cite{DeRocco:2018jwe,Obata:2018vvr,Liu:2018icu}.
In figure~\ref{figComp1}, we show the projections
for the ADBC experiment~\cite{Liu:2018icu}, taking the 2 meter cavity version
for commonality with the other meter-scale experiments 
on the plot. They assume a circulating optical power
of $10 {\rm \, kW}$, giving a stored energy of $\sim 10^{-4} \Joule$. This is many orders of magnitude below the cavity energies
of the SRF experiments, and the shot-noise-limited sensitivity
shown in the plot is correspondingly many orders
of magnitude worse. The ADBC projection does not consider
experimental issues such as polarizer efficiency, 
vibrational noise, etc, so the sensitivity of a realistic
experiment may be worse. Of course, optical experiments
may also be easier to easier and/or cheaper to 
implement than SRF experiments.


\section{Conclusions}

The nominal SRF experiments we considered above have signal 
power limited by the maximum magnetic field at the cavity
walls. Apart from increasing the cavity size,
one possibility for improving this is to use a different
superconducting material for the walls.
For example, ${\rm Nb}_3{\rm Sn}$ has $H_{sh} \simeq 0.41 \Tesla$,
and has been extensively investigated
as a potential alternative to niobium~\cite{Posen:2017ltl,padamsee1998rf}. While existing fabrication methods
lead to worse high-field performance
than niobium (potentially due to defects),
research into improving these is ongoing~\cite{Posen:2018zjb}.
Another potential benefit of alternative materials
is that they can have higher $T_c$ (e.g. $T_c \simeq 18 \kelvin$
for ${\rm Nb}_3{\rm Sn}$~\cite{padamsee1998rf});
this can help by suppressing $R_{\rm BCS}$, 
or by allowing operation at higher temperatures,
where cooling systems are more efficient.

If the thermal noise limited sensitivity projections
can actually be achieved, then further progress (for
the same cavity geometry) would depend on
`quantum engineering'.
Preparing the target
mode in a non-classical state, such as a squeezed state
or a Fock state, can improve sensitivity.
At microwave frequencies, such technologies
are being developed as part of the $\gtrsim \GHz$
axion detection program. For example, squeezed state
injection is being incorporated into Phase II of the
HAYSTAC experiment~\cite{Droster:2019fur}.

Looking beyond axion DM detection experiments, the 
more complicated cavity geometries we have discussed
 may be useful in other situations where large,
 high-frequency magnetic fields are required.
For example, microwave frequency light-through-wall
experiments to search for hidden sector particles, independent
of whether they are DM, have been proposed~\cite{Graham:2014sha,Janish:2019dpr}, and a dark photon experiment of this form is being 
constructed at Fermilab~\cite{darksrf}.
The signal strength in these experiments depends on how
large a field amplitude can be sustained in the
driven cavity.
Understanding whether or not high-field cavities can be constructed with the correct
geometry to source hidden-sector particles would require
further work.

Coming back to the prospects for low-frequency axion
DM detection, a pathfinder SRF experiment that covers
significant ALP parameter space, along the lines
of our nominal 60L cavity, seems plausibly feasible.
In principle, larger volumes and/or more advanced
cavity geometries could allow for QCD axion
sensitivity, at least insofar as fundamental noise
limits are concerned. Further work would be required to
understand whether other backgrounds,
such as drive leakage, could be mitigated.
As an alternative to static background field approaches,
SRF up-conversion would present very different experimental
challenges, and may be worth further investigation.
Unlike SRF experiments aimed at $\sim \GHz$
frequency axions~\cite{Sikivie:2010fa}, up-conversion 
of low-frequency axions
does possess parametrically better scaling than 
static-field approaches, though (as we have reviewed)
there are many compensating difficulties.
To draw an analogy, resonant bar detectors and laser
interferometers represented very different approaches
to gravitational wave detection. While static-field
experiments may well prove to be more practical,
it is important to understand the alternatives,
especially as technology (such as superconducting materials)
evolves.

Similar experimental concepts to those we
have discussed are also proposed in~\cite{slac}, which
we became aware of while this work was in progress.



\begin{acknowledgments}
	We thank Saptarshi Chaudhuri, William DeRocco, Sebastian Ellis, Peter Graham, Junwu Huang, Kent Irwin, Emilio Nanni, Chris Nantista, Raffaele Tito D'Agnolo, Sami Tantawi, Jesse Thaler, and Paul Wellander for useful discussions.
	This research was supported by the Munich Institute for Astro- and Particle Physics (MIAPP) which is funded by the Deutsche Forschungsgemeinschaft (DFG, German Research Foundation) under Germany's Excellence Strategy – EXC-2094 – 390783311.
\end{acknowledgments}


\appendix

\section{Cavity wall fields}
\label{apwall}

The Maxwell SET is traceless, with $T^\mu_\mu = 0$, so
$T^{00} = T^{ii}$. Consequently, the total energy
inside a cavity is 
\begin{equation}
	U = \int dV  T^{00} = \int dV T^{ii}
\end{equation}
We can use the standard trick~\cite{Weinberg:1972kfs,Hui:2010dn}
of rewriting $T^{ij}$ as follows,
\begin{align}
	T^{ij} &= \partial_k \left(x^i T^{kj}\right) - x^i \partial_k T^{kj} \\
	&= \partial_k \left (x^i T^{kj}\right) + x^i \partial_t T^{0j}
\end{align}
where we have used conservation of the SET, $\partial_\mu T^{\mu\nu} = 0$.
Then, time-averaging over an oscillation period of the fields
inside the cavity, 
\begin{equation}
	U = \int dV T^{ii} = \left\langle \oint_{\partial V} dS_k x^i T^{k i} 
	\right\rangle
\end{equation}
Using the boundary conditions at the conducting walls,
$n \times E = 0$, $n \cdot B = 0$, and the form of the Maxwell SET,
we have
\begin{equation}
	n_k T^{i k} = - n_k \left( E_i E_k + B_i B_k - \frac{1}{2} (E^2 + B^2)\delta_{ik})
	\right)
\end{equation}
\begin{equation}
	= \frac{1}{2} \left(B^2 - E^2\right) n_i
\end{equation}
where $n$ is the outward-pointing normal to the cavity wall.
Thus,
\begin{equation}
	U = \frac{1}{2}\left\langle \oint_{\partial V} dA \, (B^2 - E^2) (x \cdot n)
	\right\rangle
	\label{eq_usurf}
\end{equation}
Physically, this is an integrated version of the Slater
formula for the energy change due to cavity wall
deformations~\cite{slater1950microwave}. If we imagine dilating the cavity
dimensions from their initial value to zero, while keeping
the wall fields fixed, we obtain equation~\ref{eq_usurf}.


\section{SNR of resonant searches}
\label{apsnr}

In the text, we emphasised that the physical temperature $T_0$
of the cavity will generally be significantly greater 
then the signal frequency $\omega_1$.
However, as discussed in~\cite{sclong,rnl},
it is possible to reduce the thermal noise
contaminating the signal, by overcoupling the signal 
mode to a `cold' detector. Here, we review 
these SNR calculations, as they apply to our setup. 

\begin{figure}
	\begin{center}
		\includegraphics[width=.8\columnwidth]{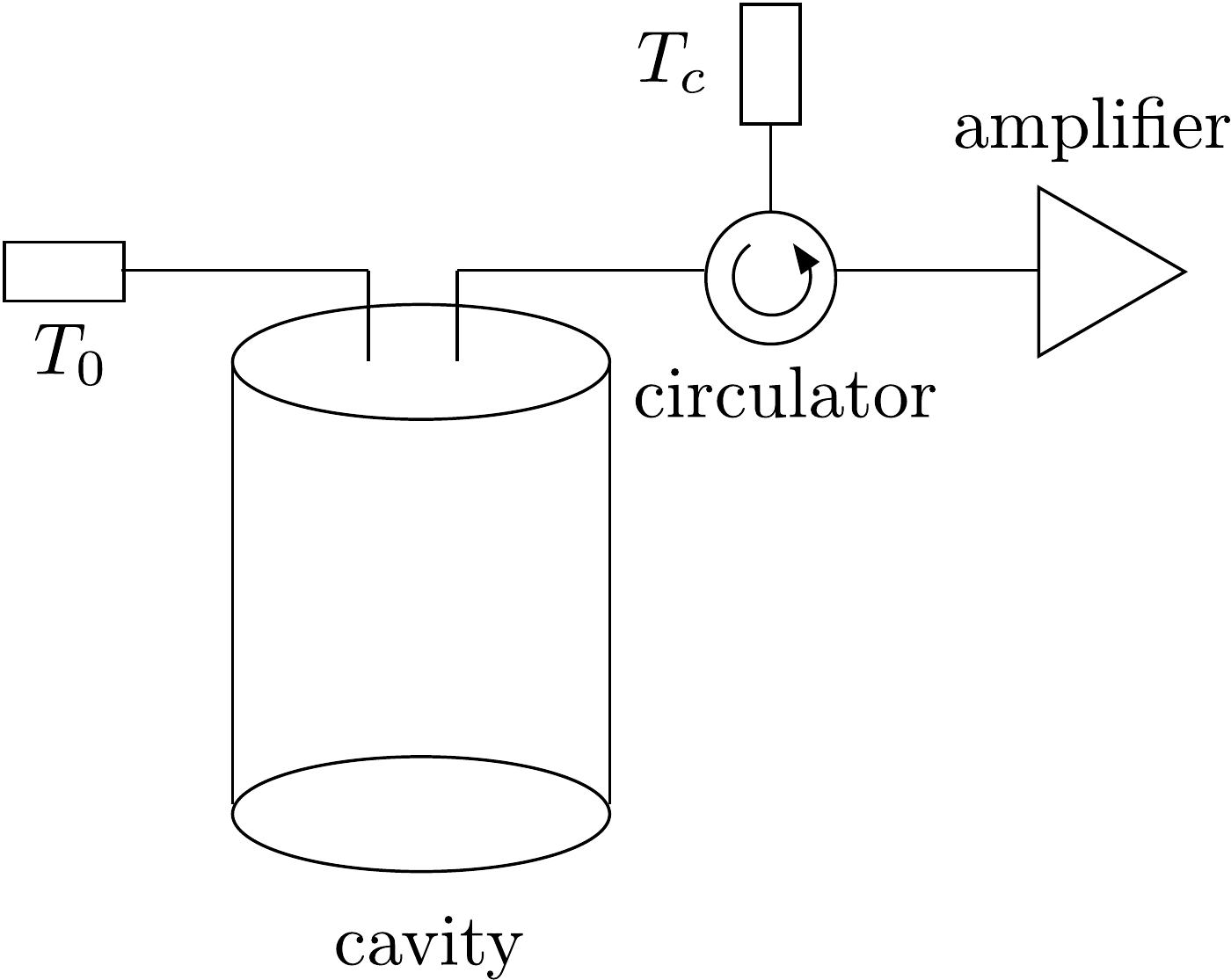}
		\caption{Signal diagram of a cavity readout system using an amplifier isolated by a circulator, as discussed in appendix~\ref{apsnr}. The
		$T_0$ load represents the thermal noise from the environment (i.e.\ the cavity walls etc), while the $T_c$ load is a cold
		load that absorbs the amplifier's back-action noise.}
	\label{figscattering}
	\end{center}
\end{figure}

Figure~\ref{figscattering} illustrates the readout
system we assume, with an amplifier isolated behind
a circulator, and a cold load absorbing the amplifier's
back-action noise.
If the output port is coupled to the signal mode
$\xi$ times more strongly than environmental dissipation
(i.e.\ a mode fluctuation loses $\xi$ times more of
its energy to the output port), then the transmission
coefficient for thermal noise from the walls is $A(\omega)
= \frac{4 \xi}{(1 + \xi)^2} \cos^2 \alpha(\omega)$, where
$\cos^2\alpha$ is as per section~\ref{secupconv}. If we assume
that the output line is impedance matched to its load,
so that no reflections are sent back to the cavity, then the
(single-sided) noise PSD at the detector input is
\begin{equation}
	S_{\rm in} = \frac{4 \xi}{(1 + \xi)^2} \cos^2 \alpha (S_{T_0} - S_{T_c}) + S_{T_c}
\end{equation}
where $T_c$ 
the effective temperature of the back-action noise
from the detector. Here, $S_T(\omega) = n_T(\omega) \omega$~\cite{Clerk_2010},
where $n_T(\omega) \equiv (e^{\omega/T} - 1)^{-1}$
is the thermal occupation number, so $S_T \simeq T$
for $T \gg \omega$.
As expected, if $T_0 = T_c$, then
the system is in equilibrium, and the noise PSD is the same
at all frequencies.

Integrating over frequencies close to the
resonance, the energy flux
from the environment to the detector is
approximately $P_{01} = (T_0 - T_c) \frac{\gamma_0 \gamma_1}{\gamma_0 + \gamma_1} = (T_0 - T_c) \gamma \frac{\xi}{(1 + \xi)^2}$,
where $\gamma_0$ and $\gamma_1$ are the damping
rates to the environment and the detector,
and $\gamma = \gamma_0 + \gamma_1$. This is
generically the result obtained
when a single harmonic oscillator interacts
with baths at different temperatures~\cite{Rieder_1967,Fogedby_2011}. The energy in the cavity mode corresponds
to an effective temperature of $T_{\rm eff}
= \frac{\gamma_0 T_0 + \gamma_1 T_c}{\gamma_0 + \gamma_1}$,
and there is an energy flux $(T_{\rm eff} - T_c) \gamma_1
= P_{01}$ to the detector port, as we would
expect
from a naive analysis.

Assuming a high-gain amplifier, the noise PSD
at the amplifier output also has contributions
from the amplifier's output noise, and 
from the amplification of vacuum
fluctuations at the input.
We can refer the amplifier's output fluctuations
to its input by dividing by the power gain $G$,
\begin{equation}
	S_n \equiv \frac{S_{\rm out}}{G} \simeq S_{\rm in} +
	S_{\rm vac} + S_{\rm amp}
\end{equation}
For a phase-insensitive amplifier, 
if the input state is coherent, then $S_{\rm vac} = 
\frac{\omega}{2}$.
For a SQL-limited amplifier, $S_{\rm amp}  =
\frac{\omega}{2}$ as well, so vacuum plus amplifier
noise combine to give `a single photon' of output
noise, in the usual phrasing~\cite{Clerk_2010}.
Below, we will write $S_a \equiv S_{\rm vac} + S_{\rm amp}$,
which for a phase-insensitive amplifier
with coherent input state is $\ge \omega$.

The PSD of absorbed power from the axion field
can be written as 
\begin{equation}
	S_{\rm abs} \simeq S_{jj} \frac{P_0}{|j_0|^2} \cos^2\alpha(\omega) \equiv S_0 \cos^2 \alpha (\omega)
\end{equation}
where $j(t) = \dot a(t) B_0(t) V_b$ (in the notation
of section~\ref{secupconv}), and $P_0$ is the power
that would be absorbed on-resonance
from a monochromatic $j(t) = j_0 \cos^2 (\omega t)$
oscillation.
For a top-hat $j$ spectrum of bandwidth
$\delta \omega_a$, we would have $S_{jj}/|j_0|^2
= 2\pi/\delta \omega_a$.
A fraction $\frac{\xi}{1 + \xi}$ of this will
enter the detector port.
Consequently, the ratio of signal to noise
PSDs, referred to the amplifier input, is 
\begin{equation}
	\frac{S_{\rm sig}}{S_n} \simeq
	\frac{\frac{\xi}{1 + \xi} \cos^2 \alpha S_0}{
		\frac{4 \xi}{(1 + \xi)^2} \cos^2 \alpha (S_{T_0} - S_{T_c})
		+ S_{T_c} + S_a}
\end{equation}

From the Dicke radiometer formula~\cite{Dicke:1946glx}, the SNR contribution
from a small frequency bin, of bandwidth
$\delta \nu$, is ${\rm SNR} \simeq \frac{S_{\rm sig}}{S_n} 
\sqrt{t_1 \delta \nu}$, where $t_1$ is the integration time. The contributions
from different frequency bins add in quadrature,
so assuming that the integration time is long enough to resolve
the spectral features of $S_{\rm sig}$ and $S_n$, we have
\begin{equation}
	{\rm SNR}^2 \simeq t_1 \int d\nu
	\left( \frac{S_{\rm sig}}{S_n} \right)^2
	\label{eqdicke1}
\end{equation}

At this point, the obvious question is what value
of $\xi$ we should select
to maximise the total SNR, in different circumstances.
To answer this, it is helpful to extract
the $\xi$ dependence from $P_0$ and $\cos^2 \alpha$; this gives
\begin{widetext}
\begin{equation}
	\frac{S_{\rm sig}}{S_n} \simeq \frac{S_{jj}}{|j_0|^2}
	\frac{\xi P_1}{
		4 \xi (S_{T_0} - S_{T_c}) + (S_{T_c} + S_a)
		\left((1 + \xi)^2 + 4 Q_1^2 \left(\frac{\omega}{\omega_1}-1\right)^2\right)
		}
		\label{eqwide}
\end{equation}
\end{widetext}
where $Q_1$ is the unloaded quality factor,
and $P_1$ is the power that would be absorbed on-resonance
in the unloaded case (c.f.\ equation~\ref{eqp1} with $Q_l = Q_1$).
Plugging this into equation~\ref{eqdicke1} reproduces
the form of equation 167 in~\cite{sclong}.

If $S_{jj}$ is even narrower than the unloaded
bandwidth of the signal mode, then to maximise sensitivity
at the axion mass we are tuned to, we simply
want to maximise $S_{\rm sig} / S_n$ on-resonance,
which is achieved at $\xi = 1$ (i.e.\ the usual
critical coupling~\cite{Clerk_2010}).
If, on the other hand, $S_{jj}$ is wide
compared to the unloaded signal mode bandwidth,
then we want to maximise
\begin{equation}
	{\rm SNR}^2 \simeq t_1 \left(\left.\frac{S_{jj} P_1}{|j_0|^2}\right|_{\omega_1}\right)^2 \int d\nu \left(\frac{\xi}{4 \xi (S_{T_0} - S_{T_c}) + \dots}\right)^2
	\label{eqsnr2xi}
\end{equation}
We write $\xi_{\rm opt}$ for the value of $\xi$ that 
maximises this integral. For example, if $T_c = T_0$,
then $\xi_{\rm opt} = 2$. 
Another case of physical interest is if $T_0 \gg T_c, S_a$.
For an SRF cavity, while cooling the cavity walls to below
$1 \kelvin$ would be prohibitively difficult,
realising a cold load at significantly lower
temperatures, and an amplifier noise temperature $\ll 1 \kelvin$,
is feasible. In this case, $\xi_{\rm opt} \simeq \frac{2 S_{T_0}}{S_{T_c} + S_a}$. The best achievable parameters
for a SQL-limited amplifier are $T_c = 0$, $S_a = \omega$,
in which case $\xi_{\rm opt} \simeq 2 T_0 / \omega_1$.
This is the optimum
overcoupling found in~\cite{sclong}.

In the $T_0 \gg T_c, S_a$ case, with $\xi = \xi_{\rm opt}$,
the loaded quality factor of the signal mode
is $Q_l \simeq Q_1 \frac{\omega_1}{2 T_0}$.
The equivalent quality factor for the expression in equation~\ref{eqwide} is $Q_s \simeq Q_l/\sqrt{3}$,
labelled the `sensitivity quality factor' in \cite{sclong}.
Physically, overcoupling
by $\xi_{\rm opt}$ reduces the loaded quality factor
of the mode, but also dilutes the thermal
noise reaching the amplifier down to 
$S_{\rm in} \simeq \frac{\omega_1}{2} \cos^2\alpha$.
Increasing $\xi$ further results in $S_a$
dominating over $S_{\rm in}$, so we reduce
the quality factor for little gain.

In~\cite{sclong}, the improvement in scan-averaged sensitivity
coming from using $\xi = \xi_{\rm opt} \simeq 2 T_0/\omega_1$ versus
$\xi \simeq 1$ is phrased as gaining sensitivity
`outside of the resonator bandwidth'. For scattering-type
setups, this applies to
the \emph{unloaded} resonator bandwidth $\sim \omega_1/Q_1$.
At $\xi = \xi_{\rm opt}$, the \emph{loaded} resonator bandwidth
(i.e.\ the physical bandwidth of mode fluctuations
in the experiment)
is, to within a $\OO(1)$ factor, the same as the sensitivity
bandwidth, as per the previous paragraph.
The improved scan-averaged sensitivity comes from this
bandwidth being parametrically larger than the unloaded
bandwidth, but the on-resonance SNR being only a $\OO(1)$ factor
smaller.
Overcoupling the signal mode reduces
the on-resonance signal power, but reduces the thermal
fluctuations in the signal mode by parametrically the same
amount, while increasing the signal mode's loaded bandwidth.

This situation can be contrasted with the case of an amplifier
used in `op-amp' mode~\cite{Clerk_2010},
which is discussed in~\cite{sclong,scshort}
for the case of flux-to-voltage amplifiers.
In the limit of very large amplifier power gain, the
power lost from the mode to the amplifier
necessarily vanishes, and the effect on the signal mode's
quality factor is very small. As discussed
in~\cite{sclong,scshort}, one can obtain an analogous increase
in scan-averaged sensitivity by `overcoupling'
to the amplifier, by the same $\xi_{\rm opt} \simeq 2 T_0 / \omega_1$
factor relative to the `critical' coupling that optimises
on-resonance sensitivity. In this case, however,
the sensitivity bandwidth is $\sim \xi_{\rm opt}$ times
larger than the \emph{physical} resonator bandwidth,
and the mode's fluctuations are larger
than those expected from the temperature $T_0$.

Returning to the scattering case, 
if we take $\xi = \xi_{\rm opt}$,
then for an axion signal with bandwidth
$\delta \omega_a \gg \omega_1 / Q_l$,
equation~\ref{eqsnr2xi} tells us that the SNR
from a single tuned configuration, with
frequency splitting within the axion 
bandwidth, is
\begin{equation}
	{\rm SNR}^2 \simeq 0.7 \frac{(P_1 / Q_1)^2 t_1 Q_a^2 Q_1 \omega_1}{T_0 T_n m_a^2} 
	\label{eqsnr2t1}
\end{equation}
where $T_n \equiv S_{T_c} + S_a$.
$P_1$ is evaluated 
for a monochromatic $j$ oscillation with $a_0$ set by the dark 
matter density, and $B_0$ set by the RMS magnetic field in the drive mode.

More generally, 
unless the axion mass is small compared to the sensitivity bandwidth,
covering an $\OO(1)$ range in axion masses requires running the
experiment in multiple different configuration --- for up-conversion,
with multiple different frequency splittings between the 
drive and signal modes. A `scan strategy' specifies which
frequency splittings to choose, and how long to stay in each of them.
In order to cover the axion mass range equally,
the frequency splittings should be spaced equally\footnote{for
$\lesssim \OO(1)$ frequency ranges,
the choice of axion masses prior, e.g.\ linear vs log-linear, does not make a significant difference.}, at frequencies differing by $\lesssim \nu_b$,
where $\nu_b = \max (\delta \nu_a, \nu_1 / Q_s)$.

If the SNR formulae derived above apply, then any such scan 
strategy will give approximately the same SNR.
The simplest example we can consider 
is to take $S_{jj}$ to be a top-hat of width $\delta \omega_a$,
and take the different frequency splittings to
be spaced equally at $\delta \omega_a$, such that
only a single configuration responds strongly at each possible axion mass.
Then, the time spent in each configuration is
$t_1 \simeq t_{\rm tot} \frac{\delta \omega_a}{\Delta m_a}$,
and the SNR from the responding configuration is 
\begin{equation}
	{\rm SNR}^2 \simeq 0.7 \frac{(P_1/Q_1)^2 t_{\rm tot} Q_a Q_1 \omega_1}{T_0 T_n m_a \Delta m_a} 
	\label{eqsnr2}
\end{equation}
For a denser set of frequency splittings, multiple
configurations will have significant response at each axion
mass. Since the absorbed power, averaged over
axion masses, is given
by equation~\ref{eqmapow}, any sufficiently dense
set of $\sim$ equal spacings will give $\sim$ the same
signal power at each axion mass, so the SNR
will still be given by equation~\ref{eqsnr2}.

However, once the time spent covering
each frequency becomes too small,
these SNR formulae become invalid.
In the case of a single configuration, once
$t_1 \lesssim Q_l / \nu_1$, we cannot resolve
the signal mode bandwidth, and the axion
signal does not have time to fully ring up the mode.

We can gain some insight into this behaviour by rewriting
equation~\ref{eqsnr2} in terms of
the average energy absorbed over the lifetime of the
experiment, $\bar W \simeq \bar P t_{\rm tot}$. This gives
\begin{equation}
	{\rm SNR}^2 
	\simeq 0.2 \frac{Q_1}{t_1 T_0 T_n} \frac{\bar W^2}{\omega_1}
	\simeq 0.4 \frac{Q_l}{\omega_1 t_1} \frac{\bar W^2}{T_n^2}
\end{equation}
As discussed in~\cite{rnl}, ${\rm SNR} / \bar W$
stops
growing for $Q_l \gtrsim t_1 \nu_1$, attaining
a maximum value of 
${\rm SNR} \simeq 0.2 \, \bar W / T_n$.


\bibliography{srf}

\end{document}